\documentclass[9pt]{olplainarticle}
\usepackage{textgreek,graphicx}
\usepackage{rotating}
\usepackage{float}
\usepackage{xr}
\usepackage{hyperref}
\usepackage[normalem]{ulem}
\usepackage{stmaryrd}
\expandafter\def\csname opt@stmaryrd.sty\endcsname
{only,shortleftarrow,shortrightarrow}
\usepackage{extpfeil}
\usepackage{ctable} 

\definecolor{mycolor}{RGB}{0, 0, 0}
\usepackage{moreverb} 

\immediate\write18{texcount -inc -incbib 
-sum main_text.tex > /tmp/wordcount.tex}

\title{Comparing antiviral strategies against COVID-19 via multiscale within-host modelling}

\author[1,2]{Farzad Fatehi}
\author[1,2,3]{Richard J Bingham} 
\author[1,2]{Eric C Dykeman}
\author[4]{Peter G Stockley}
\author[1,2,3,*]{Reidun Twarock}

\affil[1]{Departments of Mathematics, University of York, York YO10 5DD, UK}
\affil[2]{York Cross-disciplinary Centre for Systems Analysis, University of York, York YO10 5DD, UK}
\affil[3]{Department of Biology, University of York, York YO10 5DD, UK}
\affil[4]{Astbury Centre for Structural Molecular Biology, Faculty of Biological Sciences, University of Leeds, Leeds LS2 9JT, UK}
\affil[*]{Corresponding author: rt507@york.ac.uk}


\begin{abstract}
Within-host models of COVID-19 infection dynamics enable the merits of different forms of antiviral therapy to be assessed in individual patients. A stochastic agent-based model of COVID-19 intracellular dynamics is introduced here, that incorporates essential steps of the viral life cycle targeted by treatment options. Integration of model predictions with an intercellular \textcolor{mycolor}{ODE} model of within-host infection dynamics, fitted to patient data, generates a generic profile of disease progression in patients that have recovered in the absence of treatment. This is contrasted with the profiles obtained after variation of model parameters pertinent to the immune response, such as effector cell and antibody proliferation rates, mimicking disease progression in immunocompromised patients. These profiles are then compared with disease progression in the presence of antiviral and convalescent plasma therapy against COVID-19 infections. The model reveals that using both therapies in combination can be very effective in reducing the length of infection, but these synergistic effects decline with a delayed treatment start. Conversely, early treatment with either therapy alone can actually increase the duration of infection, with infectious virions still present after the decline of other markers of infection.  This suggests that usage of these treatments should remain carefully controlled in a clinical environment.
\end{abstract}

\begin{document}

\maketitle
\section{Introduction}

COVID-19 is a recently emerging infectious disease caused by severe acute respiratory syndrome coronavirus 2 (SARS-CoV-2) \cite{Lai2020}. After outbreaks of severe acute respiratory syndrome coronavirus (SARS-CoV) in 2002, and of Middle East respiratory syndrome coronavirus (MERS-CoV) in 2012, this is the third outbreak of a coronavirus since the turn of the century. Mathematical models of COVID-19 transmission at the population level have been instrumental in controlling the spread of the virus, but a detailed understanding of within-host infection dynamics is still lacking. Like SARS-CoV and MERS-CoV, SARS-CoV-2 is a betacoronavirus  in Group IV of the Baltimore classification of viruses. Compared with other single-stranded (ss)RNA viruses, coronaviruses have the longest genomes. The smallest RNA viruses, for example, are only $\sim$ 1-4kb in length, and HIV has two copies of a $\sim 10$kb genome, whilst the SARS-CoV-2 genome is a positive-sense, ssRNA molecule of $\sim$ 30kb.  As a consequence, the viral life cycle of coronaviruses is distinct from most other ssRNA viruses, and existing intracellular infection models, e.g. for hepatitis C virus \cite{Aunins2018}, cannot be applied here. Therefore, we introduce here a novel intracellular model of SARS-CoV-2 infection, that incorporates essential steps specific to coronaviral life cycles. This model enables us to study in detail the viral dynamics inside an infected cell, and provides a framework to study the impacts of antiviral treatments on the infection dynamics within an infected cell. In particular, it enables us to quantify the impact of different treatment options on the viral load that is secreted from an infected cell. 

Outcomes from the intracellular model are then integrated into an intercellular model, that takes the impact of the immune response on infection dynamics within an infected individual into account. The model has been parameterised with data from 12 patients from a study in Singapore \cite{Young2020}, enabling us to generate a generic profile of disease progression in patients that have recovered from the disease. Model predictions agree well with experimentally and clinically measured parameters such as the duration of the  \textit{incubation period}, suggesting that this scenario is representative of disease progression seen in COVID-19 patients. We then use this model to study the infection dynamics in patients with different levels of immune responses by varying parameters associated with the immune response, such as the proliferation rates of effector cells and antibodies, and the rate by which effector cells remove infected cells. Comparison of different scenarios is based on tissue damage and viral load, highlighting the impact(s) of antibodies and adaptive cell-mediated immune response on infection dynamics. 

This provides a framework in which to compare the impacts of different forms of antiviral therapy and assess  their synergies. We focus here on two prominent forms of therapy against COVID-19: remdesivir, that inhibits virus production within an infected cell \cite{Beigel2020}, and convalescent plasma (CP) therapy, whereby CP derived from recently recovered donors is transfused to the patients as an additional support \cite{Duan2020}. \textcolor{mycolor}{Recent studies have concluded that remdesivir is an effective antiviral treatment option for COVID-19 \cite{Beigel2020}. However, the rapid spread and novel nature of the disease make the detailed evaluation of effective treatment protocols difficult. Using mathematical models enables us to study in detail the effect of the drug remdesivir on viral load in a COVID-19 infection. Gon{\c{c}}alves et al. used a ``target-cell limited'' model to evaluate the efficacy of different treatment options against SARS-CoV-2 infections \cite{Gonccalves2020}. They showed that if drugs such as remdesivir are administered very early, this may help control viral load, but may not have a major effect in severely ill patients. Iwanami et al. also introduced a mathematical model to describe the within-host viral dynamics of SARS-CoV-2 and demonstrated that late timing of treatment initiation can mask the effect of antivirals in clinical studies of COVID-19 \cite{Iwanami2020}. However, none of these models have included the impact of the immune response directly into their model, which plays an important role in the outcomes of the infection. In order to analyse different aspects of viral dynamics, studying the interactions between viruses and the immune system of the host is crucial \cite{Nowak2000,Andrew2007}. In this work we take the impact of the immune response on infection dynamics into account to perform a more robust analysis of different treatment options. Regarding the CP therapy several studies performed in various countries have shown that this treatment is effective against COVID-19 infections and its safety has been well established in a randomized clinical trial (RCT) on a large population \cite{Duan2020,Joyner2020a,Joyner2020b,Ahn2020,Ye2020,Gharbharan2020,Khulood2020}. However, finding the optimal dose and time for CP therapy is still debated \cite{Duan2020}. Our intercellular model provides insights into the effects of different dosages and treatment starts in terms of infection-related quantities for CP therapy and this supports efforts in combating the COVID-19 pandemic.}

\section{Results}

\subsection{An {\emph{in silico}} model of intracellular SARS-CoV-2 infection dynamics} 

\begin{figure}[H]
	\centering
	\includegraphics[width=\linewidth]{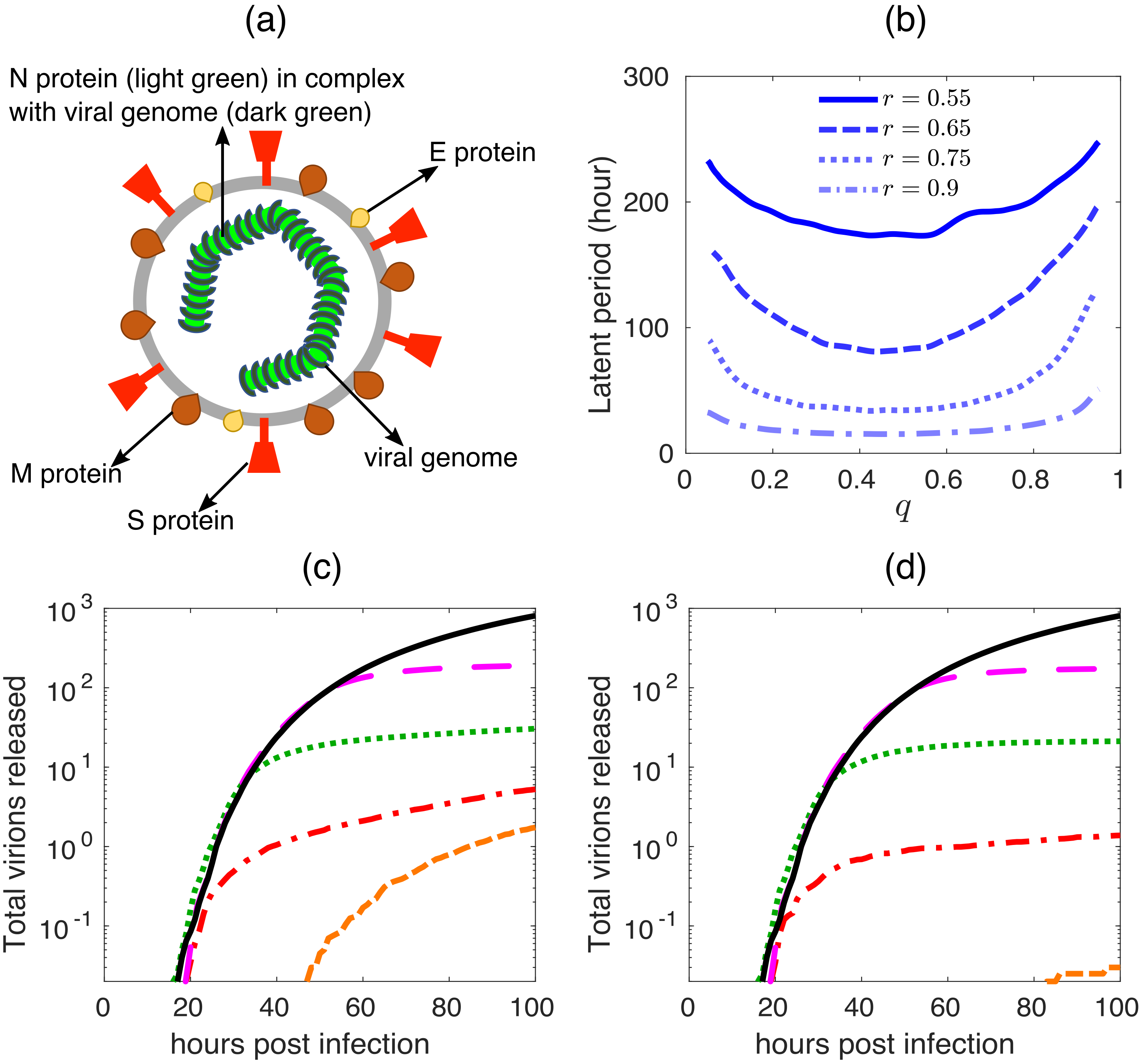}
	\caption{An earlier treatment start, especially during the latent period, is more effective. (a) Illustration of a SARS-CoV-2 virion; the viral genome (dark green) is in complex with nucleocapsid (N) protein (light green) and is enclosed by the viral envelope that is studded by other structural glycoproteins, the spike (S) protein (red), the  membrane (M) protein (maroon), and the envelope (E) protein (yellow). (b) Time lag before the release of the first virion from an infected cell; the maximal release of virions occurs when the RTC elongation probability, $r$, is high and the frameshifting rate, $q$, is between 0.3 and 0.5. (c) and (d) Profiles of viral load from an infected cell after introducing treatment at different times post infection, for a concentration of 25 (c) and 50 (d) molecules of remdesivir, respectively.  \textcolor{mycolor}{The black solid curve indicates the drug-free control. The magenta (long-dashed), green (dotted), red (dashed-dotted), and orange (dashed) curves correspond to a treatment start at 50, 30, 20, and 10 hours post infection, respectively.} Parameter values are given in (SI Table S1).}
	\label{qrtime}
\end{figure}

Our stochastic model of viral infection dynamics within an infected host cell tracks  the different viral and cellular components required for formation of progeny virus. These include the structural proteins that make up the virus: the envelope (E) protein, the membrane (M) protein, and the spike (S) protein, as well as the nucleocapsid (N) protein (Fig. \ref{qrtime}a). N protein forms a complex with the genomic RNA (gRNA), and thus aids its compaction for ease of packaging within the viral envelope. The S protein binds the \textcolor{mycolor}{Angiotensin-converting enzyme 2 (ACE-2)} receptor on human cells and is therefore essential for cell entry. The model also includes non-structural proteins that are important for the viral life cycle, such as the replicase–transcriptase complex (RTC), and keeps track of the numbers of gRNAs (and subgenomic sgRNAs) at different stages of the replication process. These include both the original  plus-sense template of the RNA molecules, as well as their negative-sense variants that arise transiently during transcription. There are nine  negative-sense sub-genomic RNAs (-sgRNAs), corresponding to different gene products. In particular, there is an individual one for each of the structural proteins, allowing the virus to produce these components in the different quantities required for formation of viral particles \cite{Bar2020}. 

The reactions modelling viral replication within the host cell are described in detail in the Supplementary Information (SI) \textcolor{mycolor}{ and {\it Materials and Methods}}. Here we provide a brief summary of the main reactions. The SARS-CoV-2 genome encodes two polyproteins, pp1a and pp1ab. The former is translated from the open reading frame ORF1a, and  the latter from the overlapping reading frames 1a and 1b \cite{Fehr2015} via a -1 ribosomal frameshift during the translational elongation step. The relative frequencies of occurrence of the ribosomal frameshift is a key mechanism of self-regulation of protein expression of the virus and hence an important parameter in our model, which is captured by the {\it frameshift probability} $q$ \cite{Nakagawa2016}.    

Proteins cleaved from pp1a and pp1ab form the RTC, which is then used by the virus for genome replication and production of the nine (-)sgRNAs. (-)sgRNAs are produced through discontinuous transcription \cite{De2016}, where elongation of nascent (-)RNA continues until the first functional transcription-regulating sequence (TRS) is encountered. A fixed proportion of RTCs will disregard the TRS motif and continue to elongate the nascent strand, while the remainder will halt synthesis of the nascent minus strand and instead synthesise (-)sgRNAs \cite{Sawicki2007}. The nine TRS motifs in the SARS-CoV-2 genome correspond to the nine sgRNAs produced, hence the choice to elongate or terminate synthesis occurs up to nine times during the elongation process \cite{Kim2020}. The (-)RNA and (-)sgRNAs produce positive-sense RNAs by recruiting RTC. The sgRNAs are translated to form proteins using cellular ribosomes. In the final step, a gRNA and the structural proteins (S, M, N, and E) form a new virion according to an assembly reaction that takes the stoichiometry of the different viral components into account \cite{Bar2020}, and the virus particle is then released from the host cell.

Stochastic simulations of the reactions were implemented using the Gillespie algorithm \cite{Gillespie1977}, and the number of particles released over the course of 100 hours was computed as the average over 200 stochastic simulations. Parameter values used (SI table S1) are predominantly based on \textcolor{mycolor}{experimentally available data \cite{Dimelow2009,Bar2020,Kim2020,Te2010,Zhang2020,Gordon2020};} for parameters for which no data were available, we ensured that our main conclusions are robust against their variation. In particular, the release rate of virions, following a time lag between infection of a host cell and its first release of viral particles, is constant and virions are secreted linearly (SI Fig. S3). 

We note, however, that the length of the time lag, and correspondingly the total number of virions released, are affected by some of these parameters and therefore warrant a more detailed investigation. For example, increasing the ribosomal protein production rate (SI Fig. S3a) or the RTC nucleotide association rate (SI Fig. S3b) decreases the time lag and therefore increases the number of virions released.  By  contrast, variation of the half-life of RTC or the formation rate of RTC from the constituent proteins does not have a significant effect on the time lag (SI Fig. S3c and d). Figure \ref{qrtime}b shows the time lag to the release of the first virion from an infected cell as a function of the ribosomal frame shifting probability $q$, and of the RTC elongation probability $r$.         

\textcolor{mycolor}{Figure \ref{qrtime}b} indicates that decreasing $r$ results in a rapid increase in the time lag: for example, when $r=0.55$ and the virus produces many sub-genomic fragments, this time lag is longer than 200 hours.  This implies that viral load is maximised for the scenario that the RTC favours continuation of the transcription process when encountering a TRS. Given the importance of frameshifting in the coronavirus life cycle to control the relative numbers of different viral components, it has been argued that the virus will have evolved to optimise this ratio. In particular, frameshift signals have been characterized experimentally previously to have efficiencies in the range of 20–45\% \cite{Baranov2005,Brierley1987,Herald1993,Plant2006}, although a recent study has suggested that the frameshift rate may be slightly higher \cite{Irigoyen2016}. Our model identifies an optimal value of $0.3<q<0.5$ as the range with the lowest time lag and hence maximal virion release (Fig. \ref{qrtime}b), in good agreement with the experimentally determined range of 20-45\%. 

\subsection{The intracellular infection model in the presence and absence of antiviral therapy}

In order to assess the impact of an antiviral drug on viral load in the context of the intracellular model, we use the example of remdesivir, which is a widely used treatment option against COVID-19. Remdesivir, originally developed as a treatment for Hepatitis C virus and later trialled for efficacy against Ebola, acts as a nucleoside analogue that mimics adenosine nucleotide \cite{Warren2016}. During the replication process, RTC may insert remdesivir molecules instead of adenosine, resulting in capping of the strand and thus terminating replication \cite{Zhang2020}. We have included additional reactions into the model that describe remdesivir binding to the RTC complexes on the gRNAs and sgRNAs to capture this (see SI for details), and track the effect of a given, fixed number of remdesivir molecules per cell on the release of viral particles from an infected host cell.

Figures \ref{qrtime}c and d show the impact on viral load released from a single infected cell for two different concentrations of remdesivir, as well as treatment starts at different times post-infection (TPIs). In Fig. \ref{qrtime}c, a free concentration equivalent to 25 remdesivir molecules per cell is considered, corresponding to a concentration of $\sim 0.06$ \textmu M (see SI) \cite{Gordon2020}.  The black solid curve indicates the viral load in the absence of treatment as a control, computed as an average over 200 stochastic simulations of the model. Magenta (long-dashed), green (dotted), red (dashed-dotted), and orange (dashed) curves show the impact(s) of treatment start at TPIs of 50, 30, 20, and 10 hours, respectively. Figure \ref{qrtime}c demonstrates that starting treatment during the latent period reduces the total number of virions released significantly. Even given a later treatment start the rate of virion production is slowed down, but the earlier treatment is started, the stronger the reduction in the virion production rate. Our results are consistent with experiments  that probed the impact of remdesivir on \textcolor{mycolor}{mouse hepatitis virus (MHV)} infection \cite{Agostini2018}, which also revealed that starting treatment earlier and during the latent period is more effective, as is the case also in other betacoronviruses. Figure \ref{qrtime}d shows the impact of doubling the drug concentration (equivalent to a concentration of 50 molecules of remdesivir per cell). In this case, starting treatment early during infection at 20 hours post infection reduced the number of virions released on average by more than half. However, starting treatment later in the infection, such as 30 or 50 hours post infection, decreases the number of released virions by a smaller fraction. This suggests that although an increased drug concentration can be beneficial, starting the treatment earlier is more effective at reducing viral load than an increase in dosage.

\textcolor{mycolor}{The intracellular model provides new insights into the release of viral particles from an infected cell, both in the absence and presence of antiviral treatment. The model shows that there is a time lag between infection of a host cell and the first release of new virions. It also shows that virions are effectively released linearly in time after the time lag which is not observed in other viral infections such as hepatitis B viral (HBV) infection \cite{Fatehi2020}. The model reveals that antiviral therapy based on remdesivir has a higher efficacy in infected cells which are in the latent period compared with those that are already producing virions. We incorporated these facts in the next section into an intercellular model of within-host infection dynamics. The model can be used as a platform for comparing different therapeutic strategies that may develop in the future against COVID-19 infections \cite{Fatehi2020}.}

\subsection{Within-host model of SARS-CoV-2 infection dynamics}

The intracellular model affords insights into the release of viral particles from an infected cell, both in the absence and presence of antiviral treatment. We integrate results from this model into an intercellular model of within-host infection dynamics in order to probe the impact of the adaptive immune response on disease progression both in the absence and presence of antiviral therapy.  Uninfected target cells ($T$) are assumed to follow logistic growth with proliferation rate $r_T$ and carrying capacity $T_m$. Inclusion of growth capacity of uninfected cells is important, because SARS-CoV-2 is detectable in patients over 20 days after the onset of symptoms \cite{Zou2020}, comparable to the time taken to regenerate the epithelium (up to 1 month \cite{Wright2001}). Uninfected cells are infected by free virions at rate $\beta$. Although infected cells in the latent phase probably die at a somewhat lower rate than productively infectious cells, we assume that all infected cells die at approximately the same rate, $\delta$, in order to minimise the number of free parameters that would complicate parameter estimation. 

\begin{figure}[H]
	\centering
	\includegraphics[width=0.6\linewidth]{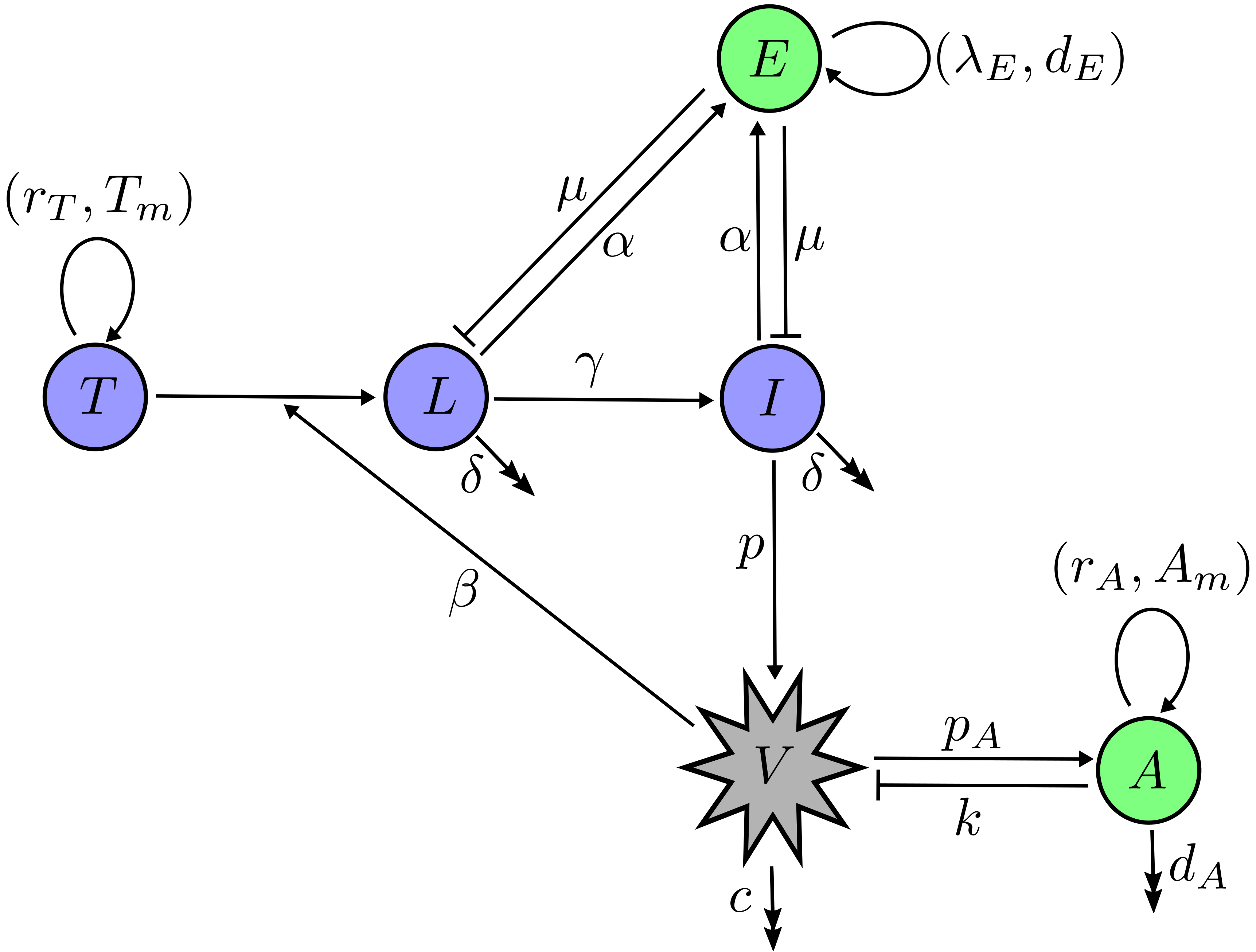}
	\caption{Diagram of the model of immune response to a viral infection. Purple circles show host cells (uninfected cells, infected cells in latent phase and productively infected cells), green circles indicated immune response (effector cells and free antibodies), and gray shows virions. Double arrow-headed lines show natural clearance. Bar-headed lines indicate the removal of infected cells and virions by immune response. Single arrow-headed lines show proliferation and production.}
	\label{diagram}
\end{figure}

Our intracellular model shows a time lag between infection of a host cell and the first release of new virions, consistent with experimental observation \cite{Bar2020}. This  effect is included into our  intercellular model via a latent phase ($L$) with a lifetime defined as $1/\gamma$, where $\gamma$ denotes the average transition rate from the latent to the productively infectious ($I$) state, i.e. when the cell sheds viral particles. The intracellular model also shows that virions are effectively released  linearly in time. Therefore we model infected cells as producing new virions $V$ at a constant production rate $p$, and assume that they are naturally cleared at rate $c$.   

Our model of the adaptive immune response consists of antibodies $A$ (humoral immune response) that remove virions at rate $k$, and effector cells $E$ (cell-mediated immune response) that kill infected cells at rate $\mu$, assuming the same rate for cells in the latent and infectious phase in order to minimise the number of free parameters. Antibodies are produced at rate $p_A$ proportional to the viral load and are degraded at rate $d_A$. After viral clearance, the antibody level is kept at a homeostatic level, because of the long-lived plasma and memory B cells. To represent this, we add a logistic term with proliferation rate $r_A$ and carrying capacity $A_m$ to the antibody equation. \textcolor{mycolor}{A fixed basal level of effector cells is assumed ($\lambda_E/d_E$), and upon infection the population of effector cells will expand at rate $\alpha(L+I)E$ \cite{Ciupe2007,Handel2010,Chenar2018}.}  $L$ and $I$ both have an impact on the immune response, because infected cells during the latent phase are producing viral proteins. Although infected cells at different stages of infection are likely to express slightly different levels of viral \textcolor{mycolor}{peptide-MHC (major histocompatibility complex)} on their surface, we assumed that the rates are the same for $L$ and $I$ in order to minimise the number of free parameters in our model.

Considering the above assumptions, the model, as illustrated in Fig. \ref{diagram}, takes on the following form:

\begin{equation}\label{inter model}
\begin{array}{l}
\vspace{0.15cm}
\displaystyle{\dfrac{dT}{dt}=r_TT(1-\dfrac{T+L+I}{T_m})-\beta TV,}\\\vspace{0.15cm}
\displaystyle{\dfrac{dL}{dt}=\beta TV-\delta L-\gamma L-\mu LE,}\\\vspace{0.15cm}
\displaystyle{\dfrac{dI}{dt}=\gamma L-\delta I-\mu IE,}\\\vspace{0.15cm}
\displaystyle{\dfrac{dV}{dt}=pI-cV-kAV,}\\\vspace{0.15cm}
\displaystyle{\dfrac{dE}{dt}=\lambda_E+\alpha(L+I)E-d_EE,}\\
\displaystyle{\dfrac{dA}{dt}=p_AV+r_AA(1-\dfrac{A}{A_m})-kAV-d_AA.}
\end{array}
\end{equation}

The model was fitted to data from 12 hospitalised patients in Singapore \cite{Young2020} using measurements of viral load (see \textit{Materials and Methods}). The parameter values derived from fitting $V$ are presented in Table \ref{patient param}. \textcolor{mycolor}{Our model captures essential features of the viral load in all patients, including the positions and heights of the first peak, and where applicable also those of the second peak (see SI Fig. S4). In all patients the viral load eventually decreases to below detectable levels, matching the clinical outcomes in these patients. We note that even details such as the slower viral decline in patients 2 and 12 are correctly represented by our model.}       

Our model predicts an \textit{incubation period}, i.e. time between infection and presentation of symptoms, of 4.25 days (3.45-5.05 95\% CI), in excellent agreement with the median SARS-CoV-2 \textit{incubation period} of roughly 5 days estimated based on clinical data elsewhere \cite{Bar2020,Lauer2020}. The average time after which antibodies appear is predicted here to be 16 days (13.9-18.1 95\% CI) after infection, again in excellent agreement with the clinically reported first detection of antibodies after 10-20 days \cite{Bar2020}. Similarly, the \textit{latent period} of 27.28 hours (26.19-28.37 95\% CI) predicted by our model agrees well with the experimentally observed latent period of 12-36 hours \cite{Harcourt2020}. 

\subsection{Immune response dynamics}

The within-host model enables the roles of different aspects of the adaptive immune response in viral clearance to be investigated in more detail. The adaptive immune response to a viral infection relies on both antibodies and effector T cells. In order to understand their respective contributions to viral clearance, we first generate a generic  progression profile based on the data from all 12 patients, and then vary parameters pertinent to different aspects of the immune response in isolation in order to probe their impact on disease progression. 

The median values from our parameter fitting were used to generate a generic  progression profile from the 12 patient data as a control (black curves in Fig. \ref{treatment inter new}a,b,c and SI Fig. S5). This control curve reveals a characteristic two-peak behaviour for viral load (Fig. \ref{treatment inter new}b), with antibodies passing the detection limit ($0.1\mbox{ ng/ml}=4\times 10^8\mbox{ molecules/ml}$ \cite{Ciupe2014}) after 14 days post infection (SI Fig. S5d). 

\begin{sidewaystable}
\footnotesize
\centering
\caption{Parameter best estimates}
\label{patient param}
\begin{tabular}{lcccccccccccc}
\hline
Patient & $\beta\times 10^{-8}$ & $\delta$ & $\gamma$ & $\mu$   & $p\times 10^{-3}$ & $k\times 10^{-10}$ & $c$   & $\alpha\times 10^{-9}$ & $p_A\times 10^{-5}$ & $r_A$ & I.P (day) & A.A (day) \\\hline
1       & 20.7                  & 0.248    & 0.9      & 0.047   & 2.18              & 2.37               & 1.24  & 17.2                   & 1.33                & 1.98  & 4 & 15  \\
2       & 2.54                  & 0.248    & 0.9      & 0.00578 & 8.13              & 3.68               & 3.15  & 1.17                   & 0.87                & 1.63  & 4 & 21  \\
3       & 9.12                  & 0.344    & 0.9      & 0.099   & 2.12              & 2.69               & 1.18  & 15.5                   & 1.41                & 2.05  & 6 & 16  \\
4       & 16.5                  & 0.2      & 1        & 0.001   & 9.4               & 2.2                & 1.61  & 1.37                   & 4.43                & 1.25  & 2 & 16  \\
5       & 5.89                  & 0.17     & 1        & 0.115   & 1.96              & 5.55               & 1.14  & 19.1                   & 1.94                & 1.37  & 6 & 23  \\
6       & 13                    & 0.275    & 0.82     & 0.187   & 3.2               & 5                  & 1.89  & 7.89                   & 3.1                 & 1.48  & 2 & 20  \\
7       & 13.3                  & 0.2      & 0.85     & 0.2     & 9.9               & 4.28               & 1.97  & 6.27                   & 6.6                 & 2.83  & 4 & 10 \\
8       & 10.4                  & 0.35     & 0.8      & 0.04    & 3.65              & 3.5                & 1.3   & 16.5                    & 7                  & 2.18  & 4 & 14  \\
9       & 10.6                  & 0.51     & 0.84     & 0.0035  & 10.4              & 2.8                & 1.1   & 0.86                   & 0.96                & 2.13  & 3 & 13  \\
10      & 12.2                  & 0.229    & 0.8      & 0.12    & 10.2              & 2.66               & 3.91  & 3                      & 5.65                & 1.57  & 6 & 16 \\
11      & 12.5                  & 0.2      & 0.9      & 0.067   & 1.3               & 4                  & 0.75  & 8.09                   & 6.5                 & 2.5   & 6 & 13  \\
12      & 6.07                  & 0.2      & 0.9      & 0.001   & 9.31              & 1.26               & 1.7   & 1.12                   & 7.7                 & 1.98  & 4 & 14  \\\specialrule{.2em}{.1em}{.1em}
median  & 11.4                  & 0.239    & 0.9     & 0.057   & 5.89               & 3.15               & 1.46  & 7.08                   & 3.77                & 1.98  & 4 & 15.5 \\
mean    & 11.1                  & 0.265    & 0.884   & 0.0739  & 5.98               & 3.33               & 1.75  & 8.17                   & 3.96                & 1.91  & 4.25 & 16 \\
std     & 4.9                   & 0.096    & 0.067    & 0.07   & 3.8                & 1.24               & 0.92  & 7.1                   & 2.64                & 0.47  & 1.4 & 3.73 \\ 
95\% CI & [8.33, 13.9] & [0.211, 0.319] & [0.846, 0.922] & [0.0343, 0.113] & [3.83, 8.13] & [2.63, 4.03] & [1.23, 2.27] & [4.15, 12.2] & [2.47, 5.45] & [1.64, 2.18] & [3.45, 5.05] & [13.9, 18.1] \\\hline 
\end{tabular}
\footnotetext{I.P: incubation period. A.A: antibodies appearance.}
\footnotetext{The units of these parameters are day$^{-1}$.}
\end{sidewaystable}
\normalsize
The cell-mediated immune response is captured in the equations by two factors: $\mu$, the removal rate of infected cells by effector cells (T cells); and $\alpha$, the proliferation rate of the effector cells. As $\lambda_E/d_E$ is the basal level of effector cells, it is assumed to be constant for each patient \cite{Ciupe2007}. Thus, $\alpha$ and $\mu$ are parameters that are likely varying in different patients. In particular, they would be expected to be lower in immunocompromised patients than for a patient with a healthy immune system \cite{Ciupe2007,Guedj2013,Conway2015,Chenar2018}. Figure \ref{treatment inter new}d (red curve; see also SI Fig. S5) demonstrates that although reduction in the value of $\mu$ can increase the damage to healthy cells ($T$) slightly, this effect is much stronger when reducing $\alpha$ (Fig. \ref{treatment inter new}g, blue curves). Even though decreasing either of these parameters causes a slower decline in viral load after the second peak (Fig. \ref{treatment inter new}e and h), a smaller value of $\alpha$ in addition increases the maximum of both peaks. Figures \ref{treatment inter new}j,k,l (magenta curves in SI Fig. S5) model the case where just the humoral immune response is weakened, i.e. where the proliferation rate of antibodies $r_A$ is reduced \cite{Ciupe2014}. In this case, viral load shows three peaks, and the damage to healthy cells recovers only slowly. This demonstrates that each component of the immune response plays a different, and crucial role in the recovery process. In particular, in immunocompromised or elderly individuals who have a weakened immune response, this could lead to significant tissue damage, with infections lasting much longer than for non-immunocompromised patients.

\subsection{The impact of different therapeutic strategies on viral dynamics in patients with different types of immune response}

\begin{figure}[H]
	\centering
	\includegraphics[width=0.86\linewidth]{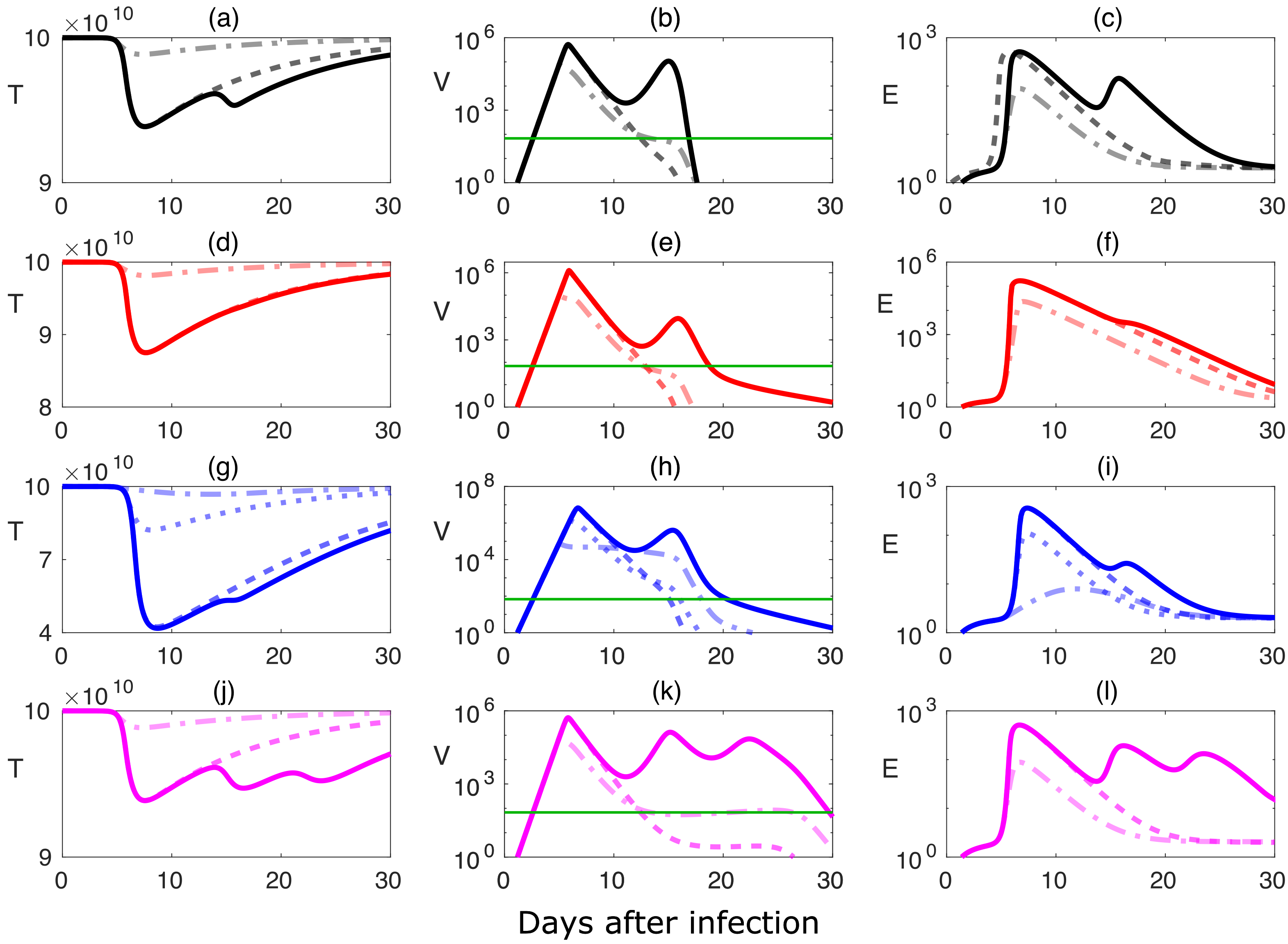}
	\caption{Antiviral treatment prevents the second viral peak and an earlier antiviral treatment start is more effective.  Solid lines indicate progression of the infection in the absence of treatment as a control. Dashed, dotted and dash-dotted curves show the result of starting treatment at 7, 6, and 5 dpi, respectively. \textcolor{mycolor}{Parameters are the median values of Table \ref{patient param} (black curves (a), (b) and (c)). Red curves ((d), (e) and (f)) correspond to the scenario of low removal rate of infected cells by effector cells ($\mu=3.5\times 10^{-4}$). Blue curves ((g), (h) and (i)) illustrate the scenario of a low proliferation rate of effector cells ($\alpha=5.4\times 10^{-10}$), and magenta curves ((j), (k) and (l)) of a low antibody proliferation rate ($r_A=1$). The green line (horizontal line in (b), (e), (h) and (k)) indicates the viral detection limit. Note that for black, red and magenta the 6 dpi scenarios have similar curves to 7 dpi, so only 5 dpi and 7 dpi are visible in the plot.}}
	\label{treatment inter new}
\end{figure}

In order to study the effects of antiviral therapy in the context of our intercellular model, we multiply the viral production rate $p$ by $(1-\epsilon)$, where $0\leq\epsilon\leq 1$ is the drug efficacy \cite{Dahari2009,Kim2012,Guedj2013,Chenar2018}. As our intracellular model (Figs. \ref{qrtime}c and d) suggests that starting treatment in the \textit{latent period} is most effective, and would effectively block the production of virions, we set $\gamma=0$ at the onset of treatment. This means that infected cells in the latent phase ($L$) do not transition to phase $I$, and begin shedding virions at a reduced rate (following a time lag $\tau$) compared with cells that were already in phase $I$ at the onset of treatment. \textcolor{mycolor}{However, as our numerical results show that the delayed model (SI Fig. S6) has the same behaviour as the model without a time delay, i.e. $\tau=0$, (Fig. \ref{treatment inter new}), we set $\tau=0$. Thus, we have the following equations for the numbers of infected cells and free virions;}
\begin{equation}\label{inter model treatment}
\begin{array}{l}
\vspace{0.13cm}
\displaystyle{\dfrac{dL}{dt}=\beta TV-\delta L-\gamma(1-\theta(t-t_R))L-\mu LE,}\\
\vspace{0.13cm}
\displaystyle{\dfrac{dI}{dt}=\gamma(1-\theta(t-t_R)) L-\delta I-\mu IE,}\\
\displaystyle{\dfrac{dV}{dt}=p(1-\epsilon)\theta(t-t_R)L+p(1-\eta\theta(t-t_R))I-cV-kAV,}\\
\end{array}
\end{equation}
where $\theta(.)$ is the Heaviside function \textcolor{mycolor}{($\theta(t)=1, \mbox{ for } t \geq 0, \mbox{ and } \theta(t)=0, \mbox{ for } t < 0 $)}, and $t_R$ is the time when the antiviral treatment (remdesivir) is introduced. \textcolor{mycolor}{Cells in phase $L$ produce virions at the reduced rate of $p(1-\epsilon)$. Our intracellular model (Fig. 1c and d) suggests that starting therapy during the \textit{latent period} can reduce the level of virions released by $\sim 99\%$ on average. We thus assume $\epsilon=0.99$ for the efficacy of remdesivir in cells in phase $L$. This is a good approximation, as all values of $\epsilon$ calculated from Fig. \ref{qrtime}c are above $98\%$, regardless of treatment start time. Cells that were already in the productively infectious phase ($I$) at the time of treatment start are assumed to produce virions at rate $p(1-\eta)$, where $\eta \leq \epsilon$ is the efficacy of remdesivir in these cells. Figure \ref{qrtime}c indicates that starting therapy during the productively infectious period can reduce the level of released virions by $\sim 90\%$ on average. Therefore we set $\eta=0.9$.  We note that these values of $\epsilon$ and $\eta$ are consistent with values used in previous models \cite{Iwanami2020,Kim2020,Gonccalves2020}}.

Figures \ref{treatment inter new}a,b,c illustrate the impact of treatment on viral clearance in a patient who would most likely recover without treatment, as the parameters have been chosen as the median values of the 12 patients who have recovered from COVID-19 without treatment (Table \ref{patient param}). In this generic scenario, a treatment start 7 days post infection (dpi), which is approximately the time of the first peak in viral load, prevents the second peak from occurring (black dashed lines in Fig. \ref{treatment inter new}). An earlier treatment start at 6 dpi also leads to the same result. However, starting treatment even earlier at 5 dpi (black dash-dot lines in Fig. \ref{treatment inter new}) also reduces the damage to healthy cells. Even though free virus declines slower in this case, the area under the viral load curve (AUC), an infection-related quantity commonly used to help the assessment of a treatment against acute viral diseases \cite{Vegvari2016}, is much smaller. In respiratory infections, even after viral clearance the immune response can cause respiratory and systemic symptoms in some incidences \cite{Vegvari2016,Ison2010}. A treatment start at 5 dpi results in a reduction in the peak of immune response cells, suggesting that early treatment perhaps could mitigate against this.

\textcolor{mycolor}{In Fig. \ref{treatment inter new} solid red (Figs. \ref{treatment inter new} d,e,f), blue (Figs. \ref{treatment inter new} g,h,i) and magenta (Figs. \ref{treatment inter new} j,k,l) curves indicate cases where different aspects of the immune response are weakened in isolation.} In particular, the solid red (Figs. \ref{treatment inter new} d,e,f) curves illustrate the case of a reduction in the removal rate of infected cells by effector cells  $\mu$ by 99\% reduction with respect to the generic case ($\mu=3.5\times 10^{-4}$), the solid blue (Figs. \ref{treatment inter new} g,h,i) curves that of a 92\% reduced proliferation rate of effector cells $\alpha$ ($\alpha=5.4\times 10^{-10}$), and the solid magenta (Figs. \ref{treatment inter new} j,k,l) curves correspond to a low antibody proliferation rate ($r_A=1$ instead of 1.98). As in the generic case above, figures \ref{treatment inter new}d-l indicate that in each case starting treatment 5 dpi reduces tissue damage, viral peak height and AUC significantly \textcolor{mycolor}{(Table S2)}, compared with treatment starts at 6 or 7 dpi, again emphasizing the importance of an early treatment start. However, early treatment increases the duration of infection compared with a later therapy start (Fig. \ref{treatment inter new}h). This suggests that viral load could persist for a longer time in such patients, who may still be infectious.

Our intercellular model also enables the modelling of drugs that operate at the level of the immune response, rather than virus production in the intracellular milieu. As an example of a therapy option of this type we study the impact of convalescent plasma (CP) therapy on viral dynamics \cite{Duan2020}. For this, we add a new variable to our model ($\tilde{A}$), which captures the antibodies that are administered as treatment. We assume that these antibodies remove virions at rate $kf$, where $0\leq f\leq 1$, implying that they are at most as efficient as the antibodies that are being developed by the body over the course of infection. The equations for the number of virions and antibodies thus have the form;

\begin{equation}\label{antibody model}
\begin{array}{l}
\vspace{0.15cm}
\displaystyle{\dfrac{dV}{dt}=pI-cV-kAV-\theta(t-t_{CP})kf\tilde{A}V,}\\\vspace{0.15cm}
\displaystyle{\dfrac{d\tilde{A}}{dt}=\theta(t-t_{CP})(-kf\tilde{A}V-d_A\tilde{A}),}
\end{array}
\end{equation}
where $\tilde{A}(t)=0$ for $t<t_{CP}$ and $\tilde{A}(t_{CP})=\tilde{A}_m$, with $\tilde{A}_m$ representing the number of antibodies per ml that are administered as treatment. $t_{CP}$ denotes the time at with the treatment is started, and $\theta(.)$ is the Heaviside function.

Our model enables the impact of CP therapy on viral dynamics to be studied for different treatment starts and doses, thus addressing the bottle-neck pointed out in the recent literature of finding the optimal dose and start for CP treatment \cite{Duan2020}. Using again median values from Table \ref{patient param} to generate a generic patient profile as a control, and using three immunocompromised cases that are presented in Fig. \ref{treatment inter new} (red, blue, and magenta curves, representing cases with reduced values for the immune response parameters $\mu$, $\alpha$, and $r_A$, respectively) we studied the impact of CP therapy. SI Fig. S7 shows the minimum level of therapeutic antibodies $\tilde{A}_m$ that is needed to reduce the AUC by 25\% and 50\% as a function of the start of treatment (in dpi) and the factor $f$ by which therapeutic antibodies are less efficient than those produced by the host during the infection. It indicates that the level of $\tilde{A}_m$ that is needed for an effective reduction in the AUC is at most around $3\times10^{11}\mbox{ molecules/ml}$. This is in good agreement with clinical data, reporting a 200 ml dose of CP and $A_m=4\times10^{12}\mbox{ molecules/ml}$ (see {\it Materials and Methods}) \cite{Duan2020,Sun2020,Ciupe2014}. Indeed, this implies that each dose would contain about $8\times10^{14}\mbox{ molecules}$, resulting in $\tilde{A}_m=2.6\times10^{11}\mbox{ molecules/ml}$ on the basis of an average level of 3 liters blood in the body \cite{Murray2015}. Hence, a reduction of the AUC by 50\% is achievable. SI Fig. S7 also shows that AUC reduction is comparable in the range $0.7\leq f\leq 0.9$, therefore we use the average value of this range ($f=0.8$) in our calculations. Our conclusions are robust for efficiencies $\geq0.15$, while for values of $f$ below 0.15, the outcomes vary for different immunocompromised cases. Using these parameters, we present a comparative analysis between antiviral and CP therapy and explore their synergistic potential.

\begin{figure}[H]
	\centering
	\includegraphics[width=\linewidth]{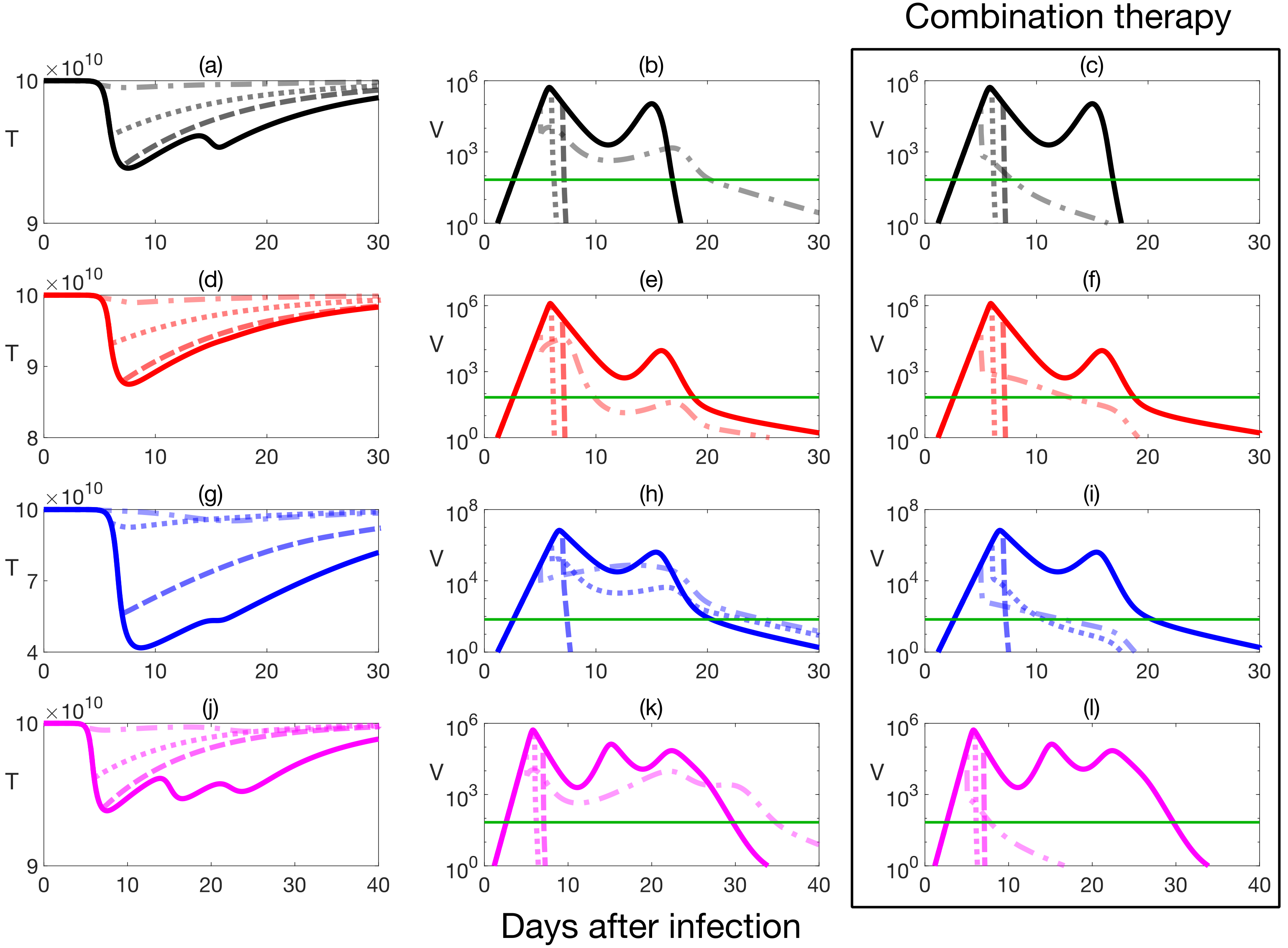}
	\caption{An early CP therapy increases the duration of infection more compared with an early antiviral therapy. Solid lines indicate progression of the infection in the absence of treatment as a control. Dashed, dotted and dash-dotted curves show the result of starting treatment at 7, 6, and 5 dpi, respectively.  Parameters are the median values of Table \ref{patient param}. \textcolor{mycolor}{Red curves ((d), (e) and (f)) correspond to the scenario of low removal rate of infected cells by effector cells ($\mu=3.5\times 10^{-4}$). Blue curves ((g), (h) and (i)) illustrate the scenario of a low proliferation rate of effector cells ($\alpha=5.4\times 10^{-10}$), and magenta curves ((j), (k) and (l)) of a low antibody proliferation rate ($r_A=1$). The green line (horizontal line in (b), (c) (e), (f), (h), (i), (k) and (l)) indicates the viral detection limit.} First and second columns indicate the impact of only CP therapy while third column show the impact of having both antiviral and CP therapy.}
	\label{antibody and rem treatment}
\end{figure}

Figure \ref{antibody and rem treatment} shows that similar to the case of antiviral therapy in Fig. \ref{treatment inter new}, an early treatment start is more effective in the reduction of tissue damage and the level of AUC compared with a later therapy start (cf. SI Fig. S8 for the equivalent of Fig. \ref{treatment inter new} for CP therapy, \textcolor{mycolor}{cf. Table S3 for AUC values}). While Fig. \ref{treatment inter new} reveals scenarios in which an antiviral therapy does not mitigate against tissue damage (such as a later treatment start at 6 or 7 dpi), Fig. \ref{antibody and rem treatment} shows that using CP therapy can reduce tissue damage even for those delayed treatment starts. Interestingly, Fig. \ref{antibody and rem treatment} shows that starting CP therapy early can increase the duration of infection more than for antiviral therapy, implying that for an early treatment starts using remdesivir is more effective. By contrast, for later treatment starts CP therapy reduces the viral load faster and decreases tissue damage compared with remdesivir therapy. Our model also enables us to probe the synergies of these treatments options. Figs. \ref{antibody and rem treatment}c, f, i, and l indicate that for an early treatment combination therapy mitigates against a longer duration of the infection \textcolor{mycolor}{(cf. Table S4 for AUC values)}. However, for later treatment starts any synergistic effects are minimal and combination therapy has the same outcome as CP therapy in isolation. Thus, unless infection is detected early, e.g. though an efficient track and trace system, treatment would likely start at a time when CP therapy in isolation would be as effective as combination therapy.

\begin{figure}[H]
	\centering
	\includegraphics[width=0.96\linewidth]{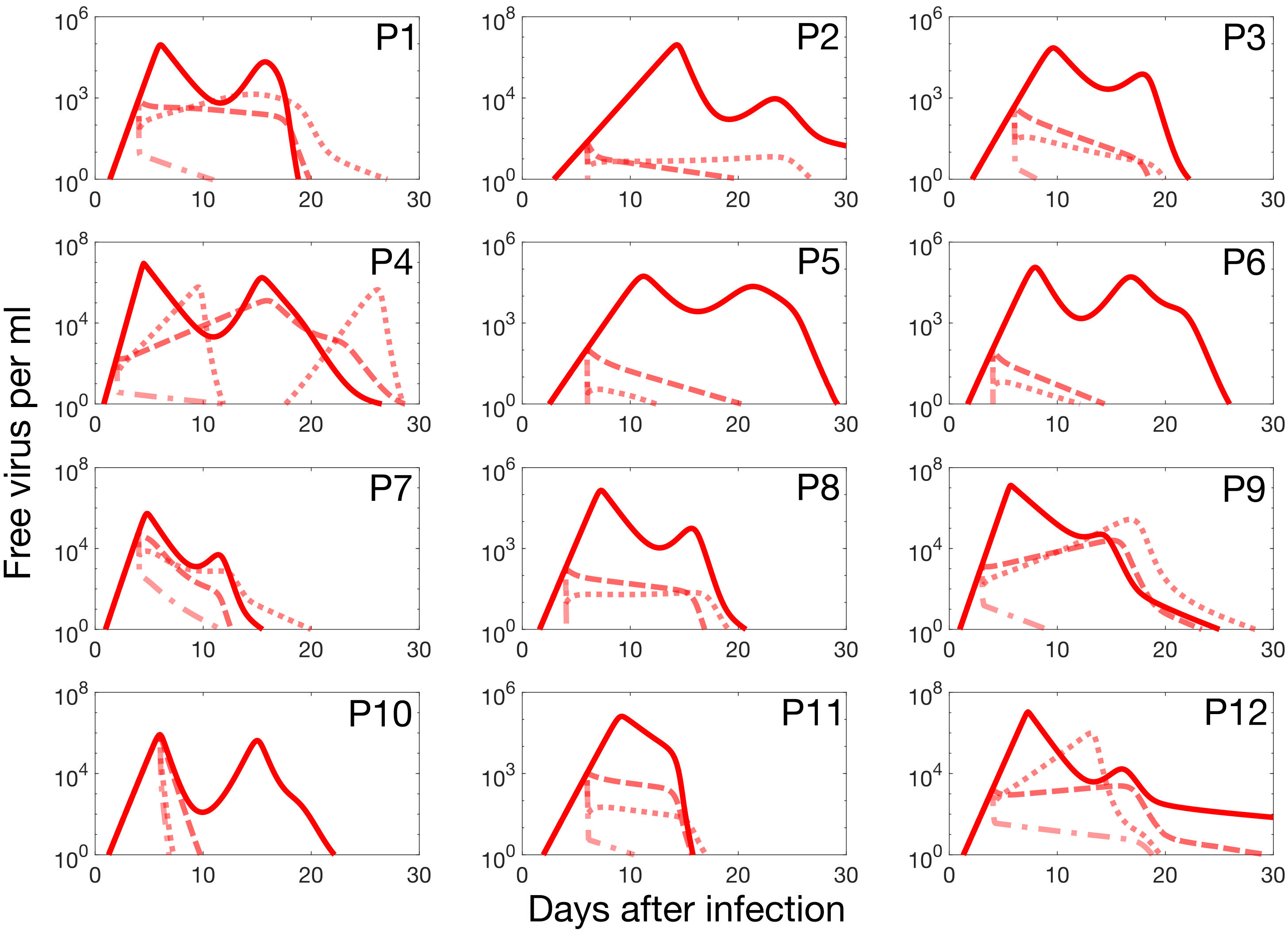}
	\caption{Starting treatment after the onset of symptoms reduces the peak in viral load and leads to faster viral clearance. Solid lines indicate the best fit to patient data. Dashed and dotted curves indicate the result of starting antiviral and CP therapy after the onset of symptoms, respectively, while dash-dotted curves indicate a combination therapy start after the onset of symptoms.}
	\label{treatment all patient}
\end{figure}

\textcolor{mycolor}{Figure \ref{treatment all patient} indicates the impact of starting treatments after the onset of symptoms, i.e. the day on which symptoms were first reported by the 12 patients in the study \cite{Young2020} used for model fitting. Both treatments reduced the duration of the infection significantly (in 67\% of the patients), enabling a faster recovery, or otherwise have no significant impact on the duration of the infection. However, in some cases in which the  peak in viral load and the AUC are significantly reduced (Table S5), the treatments have not decreased the duration of infection. This figure also shows that there are cases for which the duration of infection is not reduced by one treatment, but would be reduced by combination therapy of both treatments. In the other cases these treatment options have more or less similar effects, although CP therapy performs slightly better, and in these cases there is not a noticeable synergistic effect.}

\section{Discussion}

The severe consequences of the COVID-19 pandemic demand a concerted interdisciplinary effort to identify novel antiviral solutions. Vaccination options against SARS-CoV-2 are actively pursued and a number of treatments are now in use in the clinics, but there are  still many open questions regarding when and how best to administer these treatments, either in isolation or in combination. Whilst modelling of disease transmission has already played a key role in informing policy makers \cite{Bertozzi2020}, models of within-host dynamics have not yet had a prominent role in combating the disease. There is a precedence of intracellular modelling for other viral diseases, such as hepatitis C virus \cite{Aunins2018}. However, such models cannot be readily transferred to coronaviral infection, as viral life cycles are very different. Here we introduce a within-host model of a SARS-CoV-2 infection that contains sufficient details specific to coronaviruses  to enable antiviral strategies against SARS-CoV-2 to be compared and to analyse their synergies. We demonstrate this via a comparative analysis of an antiviral treatment (remdesivir) and CP therapy, which apart from steriod treatment are the most prominent  forms of therapy currently used against COVID-19 infections. In particular, we compare disease progression for different treatment starts and dosages, and thus provide new insights into these therapeutic options. 

\textcolor{mycolor}{Our analysis highlights, as expected and previously observed \cite{Iwanami2020,Gonccalves2020}, an early treatment start before the first peak in viral load can reduce both tissue damage and the peak viral load, especially when using a combination of both therapies. However, those models do not capture the impact of early treatment on the duration of the infection. Surprisingly, our model suggests that early treatment by either therapy alone can actually increase the duration of infection compared with a later therapy start, likely because suppressing virus production results in a reduced immune response. This implies that even though early treatment accelerates the recovery process and reduces the peak in viral load, the infection may persist for a longer time than later treatment, meaning that these patients may possibly still be infectious.}

Our model has provided insights into disease progression for different doses and treatment starts for the CP therapy \cite{Duan2020}. In particular, it enabled us to address a question recently raised in the literature as to the impact of dose and time of treatment on disease progression under CP treatment \cite{Duan2020}. Our model also enabled us to perform a comparison between the antiviral treatment and CP therapy, and explore their potential synergistic effects. The model reveals that early into the infection an antiviral treatment using remdesivir could be more effective than CP therapy, and a combination therapy can significantly reduce the duration of infection.  However, for later treatment starts, CP therapy appears to be more beneficial than antiviral therapy, and there are no longer any significant synergistic effects that would warrant combination therapy.  These insights from our within-host model suggest that the time course of infection should be considered when deciding on appropriate theraputic response to COVID-19 infection.   

\section{Materials and Methods}
\color{mycolor}
\subsection{Intracellular modelling of SARS-CoV-2 infection}

The first step in the viral lifecycle is the production of two polyproteins (pp1a and pp1ab) using the host cell ribosomes. The kinetics of ribosomes {\it in vivo} are studied using insights from a detailed stochastic model \cite{Dykeman2020}. For synthesis of the polyproteins pp1a and pp1ab, host ribosomes (denoted by R in Fig. S1b) reversibly bind to (+)RNA with binding/unbinding rates $r_{on}$ and $r_{off}$:

\begin{align*}
	&\mbox{(+)RNA}+\mbox{R}\xtofrom[r_{off}]{r_{on}}\mbox{R:(+)RNA}.
\end{align*}

We model the kinetic steps involved in ribosome initiation and transition to the elongation state (Ri$_{1a}$:(+)RNA) that occur subsequent to ribosomal binding to the (+)RNA to produce pp1a as a single kinetic step with rate $r_{in}$.

\begin{align*}
	&\mbox{R:(+)RNA}\xrightarrow{r_{in}}\mbox{Ri$_{1a}$:(+)RNA}.
\end{align*}

The ribosome then translates the pp1a gene (ORF1a) at rate $t_{1a}$. After translation of the pp1a gene, the ribosome can either frameshift -1 nt to the ORF1b reading frame, translating the polyprotein pp1ab, or terminate, releasing the polyprotein pp1a \cite{Nakagawa2016}. We model the -1 ribosomal frameshift as a reaction with rate $q\times t_f$ and the termination at ORF1a as a reaction with rate $(1-q)t_f$. If the ribosome successfully frameshifts, it completes translation of the ORF1b reading frame with rate $t_{1b}$, terminates, and releases the polyprotein pp1ab.

\begin{align*}
	&\mbox{Ri$_{1a}$:(+)RNA}\xrightarrow{t_{1a}}\mbox{R$_{1a}$:(+)RNA},\\
	&\mbox{R$_{1a}$:(+)RNA}\xrightarrow{(1-q)t_f}\mbox{pp1a}+\mbox{R}+\mbox{(+)RNA},\\
	&\mbox{R$_{1a}$:(+)RNA}\xrightarrow{q\times t_f}\mbox{Ri$_{1b}$:(+)RNA},\\
	&\mbox{Ri$_{1b}$:(+)RNA}\xrightarrow{t_{1b}}\mbox{pp1ab}+\mbox{R}+\mbox{(+)RNA}.
\end{align*}

The polyproteins pp1a and pp1ab form RTC at rate $f_{rt}$.

\begin{align*}
	&\mbox{pp1a}+\mbox{pp1ab}\xrightarrow{f_{rt}}\mbox{RTC}.\\
\end{align*}

The transcription of gRNA which leads to the formation of -gRNA and nine -sgRNAs is modelled as illustrated in Figure S2. RTC (denoted by RTC in Fig. S2) binds to the genome with binding/unbinding rates $rt_{on}$ and $rt_{off}$.

\begin{align*}
	\mbox{(+)RNA}+\mbox{RTC}\xtofrom[rt_{off}]{rt_{on}}\mbox{RT$_0$:(+)RNA},
\end{align*}

The full length genome contains functional transcription-regulating sequence (TRS) motifs which are found at the $3'$ end of the leader (leader TRS) and in front of each of the 9 ORFs (Fig. S1a) \cite{Sawicki2007}. During transcription of the full length minus strand by the RTC, the process can terminate at one of these TRS motifs, resulting in one of the 9 negative sgRNA being produced. In our model, when an RTC encounters TRS motif number $k$, it will continue the elongation of the negative strand with rate $r\times t_c$, and terminate with rate $(1-r)t_c$, resulting in the production of (-)sgRNA$_k$.

\begin{align*}
	&\mbox{RT$_0$:(+)RNA}\xrightarrow{rt_{in}}\mbox{RTi$_1$:(+)RNA}, \\
	&\mbox{RTi$_k$:(+)RNA}\xrightarrow{tr_k}\mbox{RT$_{k}$:(+)RNA},\\
	&\mbox{RT$_k$:(+)RNA}\xrightarrow{(1-r)t_c}\mbox{(-)sgRNA$_{k}$}+\mbox{RTC}+\mbox{(+)RNA},\\
	&\mbox{RT$_k$:(+)RNA}\xrightarrow{r\times t_c}\mbox{RTi$_{k+1}$:(+)RNA},\\
	&\mbox{RTi$_{10}$:(+)RNA}\xrightarrow{t_{10}}\mbox{(-)RNA}+\mbox{RTC}+\mbox{(+)RNA}.
\end{align*}

Here, $tr_k$ is the rate of RTC transcription between the TRS at site $k-1$ and site $k$ (Fig. S2) and each $\mbox{(-)sgRNA}_{k}$ for $k=1,2,...,9$, corresponds to sgRNAs for N, 8, 7b, 7a, 6, M, E, 3a, S (-)sgRNAs, respectively, whereas the $k=10$ state with transcription rate $t_{10}$ denotes the rate to transcribe the remaining length of 22kb of RNA upstream of the structural genes. This last step is responsible for the creation of full length (-)RNA. Then (-)sgRNAs and (-)RNA serve as templates for (+)sgRNAs and viral genome synthesis, respectively. The negative RNAs bind to RTC and produce positive RNAs (SI, equation S1).

(+)sgRNA$_1$, (+)sgRNA$_6$, (+)sgRNA$_7$, and (+)sgRNA$_9$ encode structural proteins N, M, E, and S, respectively, which are involved in new virion formation \cite{Kim2020}. We assume that free ribosomes are at an equilibrium level, where +sgRNAs are saturated with available ribosomes and produce protein at constant rates $t_n$, $t_m$ etc.

\begin{align*}
	&\mbox{(+)sgRNA$_1$}\xrightarrow{t_n}\mbox{(+)sgRNA$_1$}+\mbox{N},\\
	&\mbox{(+)sgRNA$_6$}\xrightarrow{t_m}\mbox{(+)sgRNA$_6$}+\mbox{M},\\
	&\mbox{(+)sgRNA$_7$}\xrightarrow{t_e}\mbox{(+)sgRNA$_7$}+\mbox{E},\\
	&\mbox{(+)sgRNA$_9$}\xrightarrow{t_s}\mbox{(+)sgRNA$_9$}+\mbox{S}.
\end{align*}

The budding of a virion is modelled as a single reaction with budding rate $k_{bud}$ as follows \cite{Aunins2018,Bar2020}: 

\begin{equation*}
	\mbox{(+)RNA}+300\mbox{S}+2000\mbox{M}+1000\mbox{N}+100\mbox{E}\xrightarrow{k_{bud}}\mbox{virion}\,.
\end{equation*}

\subsection{Modelling of antiviral strategy}

Remdesivir acts as a nucleotide analogue that mimics the adenosine structure \cite{Warren2016}. During the replication process RTC may insert remdesivir molecules rather than adenine, which caps the strand and stops the replication process at rate $r_{term}$ \cite{Zhang2020}. In order to model the impact of this drug, we assume that complexes with RTC in our model can bind (and subsequently unbind) to remdesivir molecules (Rem). Thus the reactions have the following form:

\begin{align*}
	&\mbox{RTi$_k$:(+)RNA}+\mbox{Rem}\xtofrom[k_{off}]{k_{on}}\mbox{Re:RTi$_k$:(+)RNA},\\
	&\mbox{Re:RTi$_k$:(+)RNA}\xrightarrow{r_{term}}\mbox{RTC},\\
	&\mbox{RTi$_{10}$:(+)RNA}+\mbox{Rem}\xtofrom[k_{off}]{k_{on}}\mbox{Re:RTi$_{10}$:(+)RNA},\\
	&\mbox{Re:RTi$_{10}$:(+)RNA}\xrightarrow{r_{term}}\mbox{RTC},\\
	&\mbox{RT$_k$:(+)RNA}+\mbox{Rem}\xtofrom[k_{off}]{k_{on}}\mbox{Re:RT$_k$:(+)RNA},\\
	&\mbox{Re:RT$_k$:(+)RNA}\xrightarrow{r_{term}}\mbox{RTC},
\end{align*}

where $k=1,2,...,9$.

\color{black}
\subsection{Patient data}

\textcolor{mycolor}{Our patient data comprise the first 18 confirmed patients who reported COVID-19 infection in Singapore \cite{Young2020}. Nasopharyngeal swabs were collected for up to 30 days since onset of symptoms.}  Five patients received lopinavir-ritonavir treatment, and in one patient viral load was detectable only twice, and these six patients were therefore excluded from the analysis. \textcolor{mycolor}{The viral loads were reported in cycle threshold (Ct) values, which is inversely proportional to the logarithm of the viral RNA copy number ($\log(V)=-0.3231\mbox{Ct}+14.11$) \cite{Zou2020}.} We converted Ct values to viral copies per ml. In model fitting, viral load values under the detection threshold were set at the detection limit (Ct=38).

\subsection{Intercellular model parameter estimation}

COVID-19 is a respiratory illness, so we assume that modelling insights from influenza models are applicable.  In influenza, at approximately 5 to 7 dpi mitoses are detected at the basal cell layer, and regeneration of the epithelium begins. Complete resolution of the epithelial takes up to 1 month \cite{Wright2001}. We therefore assume that the maximum proliferation rate for uninfected cells is small, and that $r_T=0.1\mbox{ day}^{-1}$. The number of host cells that express ACE-2 and \textcolor{mycolor}{transmembrane serine protease (TMPRSS)} is approximately equal to $10^{11}$ ($T_m=10^{11}$) \cite{Bar2020}, and we use $T(0)=T_m$. \textcolor{mycolor}{As SARS-CoV-2 is a novel infection, we assume that $E(0)=0$ and that the basal level of effector cells is low ($\lambda_E=1$ and $d_E=0.5\mbox{ day}^{-1}$) \cite{Ciupe2006}. However, considering a higher basal level does not change model outcomes regarding the viral dynamics as $\alpha$ and $\mu$, the proliferation rates of effector cells and removal rate of infected cells by effector cells, are estimated using viral load data fitting. Note that increasing $\lambda$ and decreasing $\mu$ simultaneously does result in the same viral dynamics, although it will change the value of the peak in effector cells. Since data is only available regarding viral load, we decided to fix the basal level of effector cells before finding other parameters \cite{Ciupe2006,Ciupe2014}.} Initially there is no specific antibody, therefore $A(0)=0$ and $d_A=0.033\mbox{ day}^{-1}$ \cite{Ciupe2014}. We use 1 \textmu g/ml \textcolor{mycolor}{immunoglobulin G (IgG)} positive control as a strong positive standard \cite{Sun2020}. Thus, we assume $A_m=1\mbox{ \textmu g/ml}=4\times 10^{12}\mbox{ molecules/ml}$. \textcolor{mycolor}{Although we are setting the individual’s antibody carrying capacity to a fixed value \cite{Ciupe2014}, we also checked that variation of the parameter does not impact the qualitative results and therefore all conclusions remain valid. $d_A$ is measured for HBV infection, but it has been shown that $r_AA(1-A/A_m)-d_AA$ is equivalent to a logistic growth of antibodies with growth rate $\rho_A=r_A-d_A$ \cite{Ciupe2014}. Since we are fitting $r_A$, fixing $d_A$ does not have a significant impact on the model. The same argument is valid for $d_E$, and as we are assuming a fixed basal level ($\lambda_E/d_E$), changing $d_E$ would not have a significant effect on our results.}

The patient data used is only available from the time after onset of symptoms, and the initial viral load at the start of the infection is  not recorded.  We therefore estimate the value $V(0)$ assuming the infection is transmitted via droplets.  The average number of expelled droplets during talking is assumed to be 1000 \cite{Loudon1967,Xie2009}. It has also been reported that more than 50\% of droplets have a size range between 50-75 \textmu m \cite{Xie2009}.  Thus, the average volume in expelled droplets during talking is equal to $1.1\times 10^{-4}\mbox{ ml}$.  The median level of viral load on the day of symptom onset in patients in this study is estimated as $5\times 10^{3}\mbox{ virion/ml}$ \cite{Ejima2020}. We assume that infected individuals infect others before the onset of symptoms, and we  therefore  assume an average level of $10^{3}\mbox{ virion/ml}$ are available for transmission.  Thus, assuming $V(0)=0.1\mbox{ virion/ml}$ appears to be a reasonable choice \cite{Gonccalves2020}. This value is also comparable to those used in modelling of influenza \cite{Pawelek2012}.

Since structural identifiability is a necessary condition for model fitting, we used the method by Castro and de Boer \cite{Castro2020} to show that our model (\ref{inter model}) is structurally identifiable (see SI Section S3 for more detail). We estimate the remaining parameters and the \textit{incubation period} (the time between the beginning of the infection and the onset of symptoms) by fitting $V$ from the model (\ref{inter model}) to patient data individually in Matlab using the method in Ciupe et al. \cite{Ciupe2014} which uses the minimum search function for data fitting. Although we fitted patient data individually, which is suboptimal compared to population fitting using mixed effects, the outcomes of the model were in agreement with clinically measured values, such as the incubation period and the onset of appearance of antibodies in the body. The resulting parameter values are presented in Table 1. Decreasing/increasing of $V(0)$ ($V(0)=0.01\mbox{ virion/ml}$ or $V(0)=1\mbox{ virion/ml}$) does not change the estimated values of parameters significantly and only changes the estimated \textit{incubation period} by $\pm 1$ day. Additionally, we used residual bootstrapping to provide 95\% confidence intervals (CIs) for the parameter estimates following \cite{Ciupe2007} (see also SI Table S2). For each set of patient data $\{V_1, V_2, ...,V_n\}$, we calculated the normalised residuals $\epsilon_i={V_i}/{\overline{V}_i}$, $i=1, 2, ...,n$, where $\{\overline{V}_1, \overline{V}_2, ..., \overline{V}_n\}$ denotes the viral load values predicted by the model. We then created the set $\{V^*_1, V^*_2, ..., V^*_n\}$ where $V^*_i=\overline{V}_i\times\epsilon_j$ for $\epsilon_j$ randomly chosen to be any of the normalised residuals or 1, the latter to include the option that the data remains unchanged. We created 50 samples and fitted each individually to the data. We then calculated the 95\% CI for each given parameter across the 50 parameter sets (SI Table S2). We generated 500 simulations based on randomly chosen  parameters from the 50 parameter sets, and then used these curves to calculate the 95\% CI for each patient (see red shaded areas in Fig. S4). As the 95\% CIs have negligible width compared with the widths of the curves, given the logarithmic scale, we also added the mean plus/minus standard deviation as shaded green areas in order to reflect the noise in the data, especially for P4 and P6. We note that the predicted two-peak behaviour is consistent with observations in W\"{o}lfel et al. \cite{Wolfel2020}, and indeed is expected in any model that includes the adaptive immune response \cite{Ke2020}.
\vspace{-0.1cm}
\section*{Acknowledgment}
RT acknowledges funding via an EPSRC Established Career Fellowship (EP/R023204/1) and a Royal Society Wolfson Fellowship (RSWF/R1/180009). RT \& PGS acknowledge support from a Joint Wellcome Trust Investigator Award (110145 \& 110146).
\vspace{-0.1cm}


\end{document}


\maketitle

\section{A stochastic model of intracellular dynamics of SARS-CoV-2}
The intracellular model contains reactions representing the translation, transcription and assembly of SARS-CoV-2 in cells. The viral entry step is not included into the model as we are focusing on the intracellular dynamics. Here we will detail the reactions underlying each process and describe the stages which have been modelled mathematically. All parameter values are given in Table \ref{param}. Stochastic simulations of the reactions were implemented using a Gillespie algorithm \cite{Gillespie1977}, and seeded with one virion at the start of the infection.

\subsection{Model for translation in SARS-CoV-2}
\begin{figure}[H]
	\centering
	\includegraphics[width=1\linewidth]{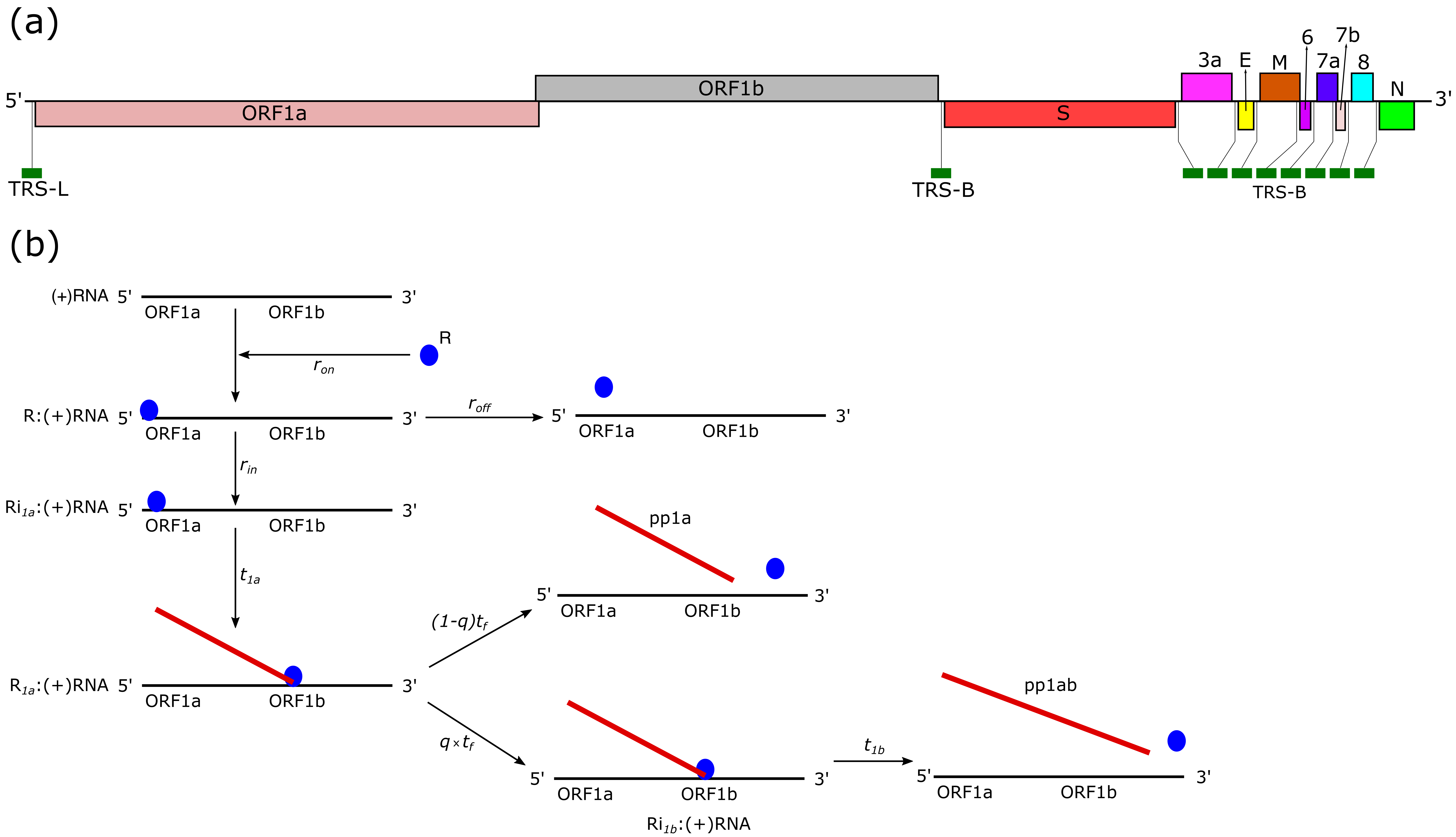}
	\caption{Schematic presentation of the SARS-CoV-2 genome and genome translation process. (a) Viral genome encodes ORF1a and ORF1b which are are translated and nine subgenomic RNAs (sgRNA) which encodes viral structural proteins (S, M, N, E) and accessory proteins. (b) The ribosome translates the pp1a gene (ORF1a). After translation of the pp1a gene, the ribosome can either frameshift -1 nt to the ORF1b reading frame, translating the polyprotein pp1ab, or terminate releasing the polyprotein pp1a. Probability of frameshifting from the ORF1a to ORF1b reading frames is $q$. }
	\label{Rib}
\end{figure}

The SARS-CoV-2 genome, illustrated in Fig. \ref{Rib}a, shows the gene products encoded by the 29,903 nts ssRNA viral genome. The replicase gene comprises roughly 22kb and encodes two polyproteins (pp1a and pp1ab) which undergo cotranslational proteolysis into the 16 non-structural proteins (nsp1-nsp16) required by the virus for infection in the host cell. The remaining genes 3' of the replicase gene encode for the structural proteins S, M, E, N. Structural and non-structural proteins are translated from different messenger RNAs that are produced by the viral replicase-transcriptase complex (RTC) from viral gRNA. The replicase gene is translated from full-length viral RNA whereas the structural genes are translated from sub-genomic mRNAs synthesised by the viral RTC \cite{Fehr2015}.

We model the translational process from the mRNAs produced by the RTC as follows (Fig. \ref{Rib}b). For synthesis of the polyproteins pp1a and pp1ab, host ribosomes (denoted by R in Fig. \ref{Rib}b) reversibly bind to with binding/unbinding rates $r_{on}$ and $r_{off}$, $K_d=0.08$ \textmu M \cite{Dimelow2009}.

\begin{align*}
	&\mbox{(+)RNA}+\mbox{R}\xtofrom[r_{off}]{r_{on}}\mbox{R:(+)RNA}.
\end{align*}

As the ribosome can bind and unbind, we assume after binding it can either unbind or start the elongation process. Thus, we model the kinetic steps involved in ribosome initiation and transition to the elongation state (Ri$_{1a}$:(+)RNA) that occur subsequent to ribosomal binding to the as a single kinetic step with rate $r_{in}$.

\begin{align*}
	&\mbox{R:(+)RNA}\xrightarrow{r_{in}}\mbox{Ri$_{1a}$:(+)RNA}.
\end{align*}

The ribosome then translates the pp1a gene (ORF1a) at rate $t_{1a}$. After translation of the pp1a gene, the ribosome can either frameshift -1 nt to the ORF1b reading frame, translating the polyprotein pp1ab, or terminate, releasing the polyprotein pp1a \cite{Nakagawa2016}. We model the -1 ribosomal frameshift as a reaction with rate $q\times t_f$ and the termination at ORF1a as a reaction with rate $(1-q)t_f$. These rates give a probability of frameshifting from the ORF1a to ORF1b reading frames of $q$. If the ribosome sucessfully frameshifts, it completes translation of the ORF1b reading frame with rate $t_{1b}$, terminates, and releases the polyprotein pp1ab. As the viral genome needs to bind to the RTC for replication, we assume that (+)RNA can have only one ribosome or one RTC on at a time. Details of the reactions are listed below.

\begin{align*}
	&\mbox{Ri$_{1a}$:(+)RNA}\xrightarrow{t_{1a}}\mbox{R$_{1a}$:(+)RNA},\\
	&\mbox{R$_{1a}$:(+)RNA}\xrightarrow{(1-q)t_f}\mbox{pp1a}+\mbox{R}+\mbox{(+)RNA},\\
	&\mbox{R$_{1a}$:(+)RNA}\xrightarrow{q\times t_f}\mbox{Ri$_{1b}$:(+)RNA},\\
	&\mbox{Ri$_{1b}$:(+)RNA}\xrightarrow{t_{1b}}\mbox{pp1ab}+\mbox{R}+\mbox{(+)RNA},
\end{align*}

The polyproteins pp1a and pp1ab are cleaved by two proteases, papainlike protease (PLpro; corresponding to nsp3) and a main protease, 3C‑like protease (3CLpro; corresponding to nsp5). Polyproteins pp1a and contains the nsps 1–11 and pp1ab contains nsps 1–10 and nsps 12-16 \cite{Fehr2015,Romano2020}. It has been proposed that nsp2-16 collectively constitute a functional RTC in infected cells \cite{Egloff2004,v2019}. These nsps have their own specific roles in the replication process, but the functions of some of them are not well understood \cite{Chen2020}. For example nsp12 is the RNA-dependent RNA polymerase (RdRp); nsp13 is the NTPase/helicase; nsp14 is a proof-reading exonuclease \cite{Fehr2015}. As the exact stoichiometry of these nsps in an RTC molecule is unknown \cite{Hagemeijer2010} and nsp11 which is a part of RTC is only in pp1a, we assume that pp1a and pp1ab have equal impact on RTC formation and this step is modelled as follows:

\begin{align*}
	&\mbox{pp1a}+\mbox{pp1ab}\xrightarrow{f_{rt}}\mbox{RTC},\\
\end{align*}

Finally, we add the natural clearance of these proteins via the reactions

\begin{align*}
	&\mbox{RTC}\xrightarrow{\delta_r}\mbox{0},\\
	&\mbox{pp1a}\xrightarrow{\delta_a}\mbox{0},\\
	&\mbox{pp1ab}\xrightarrow{\delta_b}\mbox{0},
\end{align*}

(+)sgRNA$_1$, (+)sgRNA$_6$, (+)sgRNA$_7$, and (+)sgRNA$_9$ encodes structural proteins N, M, E, and S, respectively, which are involved in new virion formation \cite{Kim2020}. The other (+)sgRNA encode accessory proteins which are hypothesised to interfere with the host innate immune response and their functions are poorly understood \cite{De2016}. Thus we only include Synthesis of S, M, N, and E proteins into our model. Furthermore, similar to previous work on intracellular modelling of HCV, where the number of free ribosomes is fitted to obtain the viral dynamics that was observed experimentally \cite{Aunins2018}, we assume that free ribosomes are at an equilibrium level, where sgRNAs are saturated with available ribosomes and produce protein at a constant rate $t_n$, $t_m$ etc.

\begin{align*}
	&\mbox{(+)sgRNA$_1$}\xrightarrow{t_n}\mbox{(+)sgRNA$_1$}+\mbox{N},\\
	&\mbox{(+)sgRNA$_6$}\xrightarrow{t_m}\mbox{(+)sgRNA$_6$}+\mbox{M},\\
	&\mbox{(+)sgRNA$_7$}\xrightarrow{t_e}\mbox{(+)sgRNA$_7$}+\mbox{E},\\
	&\mbox{(+)sgRNA$_9$}\xrightarrow{t_s}\mbox{(+)sgRNA$_9$}+\mbox{S},\\
	&\mbox{N}\xrightarrow{\delta_n}0,\\
	&\mbox{M}\xrightarrow{\delta_m}0,\\
	&\mbox{E}\xrightarrow{\delta_e}0,\\
	&\mbox{S}\xrightarrow{\delta_s}0.
\end{align*}

The E protein is found in small quantities within the virion. We assume each SARS-CoV-2 virion has 100 copies of E protein assembled into 20 pentameric structures, and 300 copies of S protein assembled into 100 trimeric structures \cite{Bar2020}. The level of M and N proteins in SARS-CoV-2 consists of approximately 2000 and 1000 copies, respectively \cite{Bar2020}. Note that in this model we assume that SARS-CoV-2 has the similar levels of E, S, M and N proteins to SARS-CoV-1 which was reported for \cite{Bar2020}. Similar to the previous intracellular models for HCV infection which have modelled the assembly and budding of virions as a single reaction \cite{Aunins2018}, we model the budding of a virion is modelled as a single reaction with budding rate $k_{bud}$ as follows: 

\begin{equation*}
	\mbox{(+)RNA}+300\mbox{S}+2000\mbox{M}+1000\mbox{N}+100\mbox{E}\xrightarrow{k_{bud}}\mbox{virion}\,.
\end{equation*}

\subsection{Model for transcription in SARS-CoV-2}
\begin{figure}[H]
	\centering
	\includegraphics[width=1\linewidth]{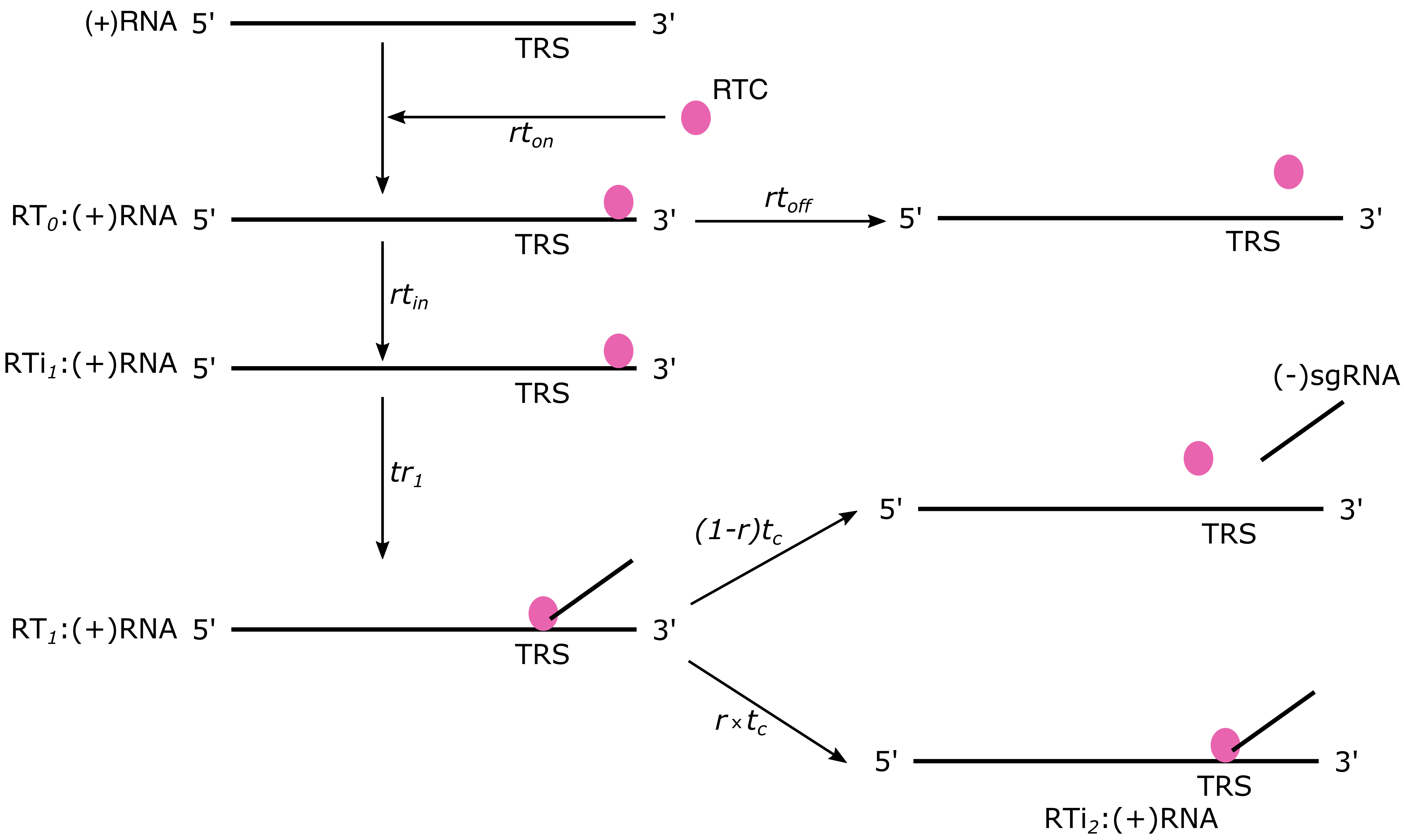}
	\caption{Transcription of SARS-CoV-2 genome via the RTC complex. During transcription process a subset of 9 sub-genomic minus strands are produced through discontinuous transcription. The full length genome contains functional (TRS) motifs which are found at the $3'$ end of the leader (leader TRS) and in front of each of the 9 ORFs (Fig. \ref{Rib}a). When an RTC encounters TRS motif, it will either continue the elongation of the negative strand or terminate the transcription, resulting in the production of (-)sgRNAs. Probability of RTC elongation is $r$.}
	\label{RTC}
\end{figure}

Transcription of the RNA genome of SARS-CoV-2 occurs via the interaction of the RTC complex with genomic (+)ssRNA, genomic (-)ssRNA, and negative sense sub-genomic fragments. During transcription of (+)ssRNA, a subset of 9 sub-genomic minus strands, which encode all structural proteins, are produced through discontinuous transcription \cite{De2016}. Each sub-genome contains a $5'$ leader sequence corresponding to the $5'$ end of the genome which allows interaction between host ribosomes and (+)sgRNAs. Each sub-genomic RNA consists of a single ORF that encodes for a structural or accessory protein \cite{Sawicki2007}. In SARS-CoV-2, the order of these ORFs (from 5' to 3') is S, 3a, E, M, 6, 7a, 7b, 8, and N \cite{Kim2020}. The full length (+)ssRNA genome contains functional transcription-regulating sequence (TRS) motifs which are found at the $3'$ end of the leader (leader TRS) and in front of each of the 9 ORFs. During transcription of the full length minus strand by the RTC, the process can terminate at one of these TRS motifs, resulting in one of the 9 negative sgRNA being produced. In our model, when an RTC encounters TRS motif number $k$, it will continue the elongation of the negative strand with rate $r\times t_c$, and terminate with rate $(1-r)t_c$, resulting in the production of (-)sgRNA$_k$ after extension by transcription of the $5'$ end of the genome \cite{Sawicki2007}. The completed minus-strand sgRNA serves as a template for (+)sgRNA synthesis. Figure \ref{RTC} illustrates the transcription reactions that we model in this work. For transcription from full length genome (+)ssRNA, RTC (denoted by RTC in Fig. \ref{RTC}) reversibly binds to 3' UTR with binding/unbinding rates $rt_{on}$ and $rt_{off}$, $K_d=0.1$\textmu M \cite{Te2010}.

\begin{align*}
	\mbox{(+)RNA}+\mbox{RTC}\xtofrom[rt_{off}]{rt_{on}}\mbox{RT$_0$:(+)RNA},
\end{align*}

We model the kinetic steps involved in RTC initiation and transition to the elongation state that occur subsequent to RTC binding to the 3' UTR as a single kinetic step with rate $rt_{in}$. The formation of (-)sgRNAs can be modelled as follows:

\begin{align*}
	&\mbox{RT$_0$:(+)RNA}\xrightarrow{rt_{in}}\mbox{RTi$_1$:(+)RNA}, \\
	&\mbox{RTi$_k$:(+)RNA}\xrightarrow{tr_k}\mbox{RT$_{k}$:(+)RNA},\\
	&\mbox{RT$_k$:(+)RNA}\xrightarrow{(1-r)t_c}\mbox{(-)sgRNA$_{k}$}+\mbox{RTC}+\mbox{(+)RNA},\\
	&\mbox{RT$_k$:(+)RNA}\xrightarrow{r\times t_c}\mbox{RTi$_{k+1}$:(+)RNA},\\
	&\mbox{RTi$_{10}$:(+)RNA}\xrightarrow{t_{10}}\mbox{(-)RNA}+\mbox{RTC}+\mbox{(+)RNA}.
\end{align*}

Here, $tr_k$ is the rate of RTC transcription between the TRS at site $k-1$ and site $k$ (Fig. \ref{RTC}) and each sub-genomic (-)RNA for $k=1,2,...,9$, corresponds to sgRNAs for N, 8, 7b, 7a, 6, M, E, 3a, S (-)sgRNAs, respectively, whereas the $k=10$ state with transcription rate $t_{10}$ denotes the rate to transcribe the remaining length of 22kb of RNA upstream of the structural genes. This last step is responsible for the creation of full length (-)genomic RNA.

To model the synthesis of the 9 (+)sgRNAs that are required for translation by the host ribosomes, we assume that RTCs can bind to negative sense RNAs producing a polysome arrangement with up to $n_k$ RTCs present on (-)sgRNA$_k$. RTCs bind to (-)sgRNA one at a time with binding rate $b_k$, until the (-)sgRNA$_k$ is saturated with $n_k$ RTCs. After saturation, the last RTC (i.e. the $n_k$th) at the 5' end of the (-)sgRNA can terminate transcription with rate $p_i$, resulting in the production of a (+)sgRNA and release of an RTC. The resulting (-)sgRNA with $n_k-1$ RTCs can bind another RTC at the 3' end. This reaction models the movement of the remaining RTCs down the RNA strand and movement of the 5' most RTC to the 5' end for termination and daughter strand release. The synthesis of (+)sgRNA is described via the following reactions:

\begin{align}
    &\mbox{(-)sgRNA$_k$}+\mbox{RTC}\xrightarrow{b_k}\mbox{(-)sgRNA$_k$:RT},\nonumber\\
    &\mbox{(-)sgRNA$_k$:jRT}+\mbox{RTC}\xrightarrow{b_k}\mbox{(-)sgRNA$_k$:(j+1)RT}, \\
	&\mbox{(-)sgRNA$_k$:$n_k$RT}\xrightarrow{p_k}\mbox{(+)sgRNA$_k$}+\mbox{(-)sgRNA$_k$:$(n_k-1)$RT}+\mbox{RTC},\nonumber
\end{align}

where $k=1,2,...,9$ denote the 9 sgRNA species and $j$ indicate the number of RTC currently bound to the (-)sgRNA. We model the replication of the full length (+)RNA from the (-)RNA using the same model as the sgRNAs and denote full length (-)RNA as $k=10$ and use binding rate $b_{10}$ with a maximal RTC number of $n_{10}$ and termination rate of $p_{10}$. Finally, we allow decay of the viral (+) and (-)sgRNA via the reactions

\begin{align*}
	&\mbox{(-)sgRNA$_k$:$n_k$RT}\xrightarrow{d_k}0,\\
	&\mbox{(+)sgRNA$_k$}\xrightarrow{d_k}0,\\
	&\mbox{(-)sgRNA$_k$}\xrightarrow{d_k}0,
\end{align*}

\subsection{Remdesivir related reactions}
Remdesivir acts as a nucleotide analogue that mimics the adenosine structure. It was originally developed as a treatment for Hepatitis C virus and later repurposed for Ebola \cite{Warren2016}. Recent studies have pointed to remdesivir as an effective antiviral treatment option for Covid-19 \cite{Beigel2020}. During the replication process RTC may insert remdesivir molecules rather than adenine, which caps the strand and stops the replication process \cite{Zhang2020}. In order to model the impact of this drug, we assume that complexes with RTC in our model can bind (and subsequently unbind) to remdesivir molecules (Rem). Thus the new reactions have the following form:

\begin{align*}
	&\mbox{RTi$_k$:(+)RNA}+\mbox{Rem}\xtofrom[k_{off}]{k_{on}}\mbox{Re:RTi$_k$:(+)RNA},\\
	&\mbox{Re:RTi$_k$:(+)RNA}\xrightarrow{r_{term}}\mbox{RTC},\\
	&\mbox{RTi$_{10}$:(+)RNA}+\mbox{Rem}\xtofrom[k_{off}]{k_{on}}\mbox{Re:RTi$_{10}$:(+)RNA},\\
	&\mbox{Re:RTi$_{10}$:(+)RNA}\xrightarrow{r_{term}}\mbox{RTC},\\
	&\mbox{RT$_k$:(+)RNA}+\mbox{Rem}\xtofrom[k_{off}]{k_{on}}\mbox{Re:RT$_k$:(+)RNA},\\
	&\mbox{Re:RT$_k$:(+)RNA}\xrightarrow{r_{term}}\mbox{RTC},\\
	&\mbox{(-)sgRNA$_l$:jRT}+\mbox{Rem}\xtofrom[k_{off}]{k_{on}}\mbox{Re:(-)sgRNA$_l$:jRT},\\
	&\mbox{(-)sgRNA$_l$:jRT}\xrightarrow{r_{term}}\mbox{RTC}+\mbox{(-)sgRNA:(j-1)RT},\\
\end{align*}

where $k=1,2,...,9$ and $l=1,2,...,10$.

\subsection{Parameters for the intracellular model}
\subsubsection{RNA and protein decay rates}
Wada et al. \cite{Wada2018} have measured half-lives of the genome and S-sgRNA for mouse hepatitis virus (MHV), a prototypic member of the CoV family belonging to the same genus (\textit{Betacoronavirus}) as SARS-CoV-2. Half-lives of S-sgRNA and genome are measured as 5.96 and 5.41 hours, respectively. Therefore, we assume that half-lives of (+)RNA, (-)RNA, and all positive and negative sgRNA are 6 hours. Thus, their decay rate is equal to $\ln2/6\mbox{ hour}^{-1}$. Half-life of E protein is 1 hour \cite{Keng2011}. We are assuming that half-lives of N, M, S, pp1a and pp1ab are also 1 hour. As half-life of the viral polymerase of HCV (like SARS-CoV-2 a +ssRNA virus) has been measured as 6 hours \cite{Aunins2018}, we assume that half-life of RTC is also 6 hours. See Table \ref{param} for full details on these constants.

\subsubsection{Ribosome translation and RTC transcription rates}
\begin{figure}[H]
	\centering
	\includegraphics[width=1\linewidth]{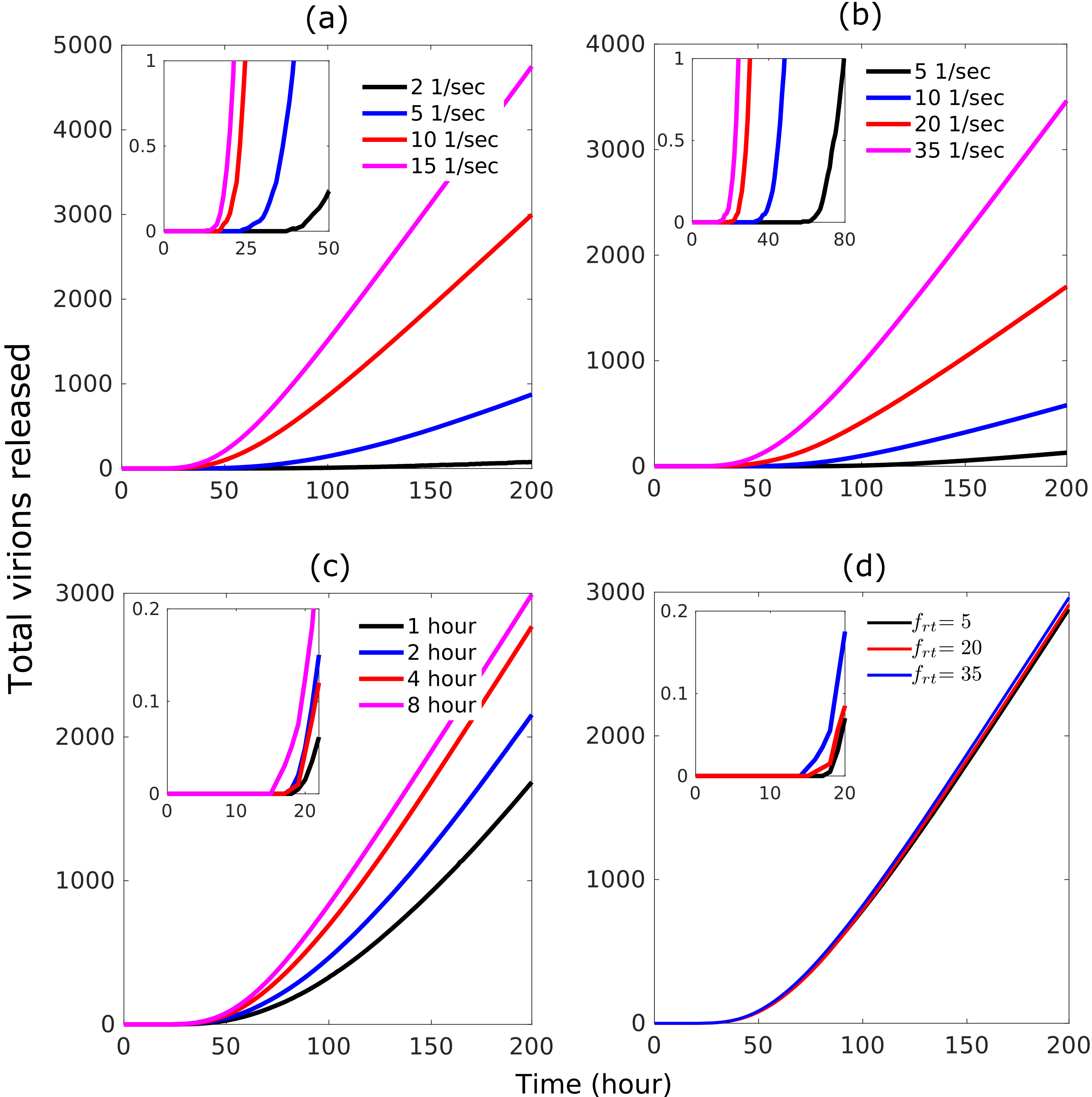}
	\caption{Total number of released virions using parameter values from Table \ref{param}. Total number of secreted virions from an infected cell as a function of (a) ribosome amino acid association rate, and (b) RTC nucleotide association rate. Total number of released viral particles as a function of (c) RTC half-life, and (d) RTC formation rate ($f_{rt}$).}
	\label{virion release}
\end{figure}

Dimelow and Wilkinson \cite{Dimelow2009} have built a detailed kinetic model of translation initiation in yeast. They have estimated that the forward rate constants for ribosome binding are in the range of 1–50 (\textmu M s)$^{-1}$. In this model we assume that the binding rate of ribosome to RNA is 25 (\textmu M s)$^{-1}$ and its unbinding rate is 2 s$^{-1}$ \cite{Dimelow2009,Wilkinson2008}. RTC binds to RNAs with dissociation constant ($K_d=\dfrac{rt_{off}}{rt_{on}}$) equal to 0.1 \textmu M \cite{Te2010}. It has been observed that increasing/decreasing RTC binding rate does not have a significant effect on the outcome as it will also increase/decrease RTC unbinding rate \cite{Fatehi2020}. Therefore we consider that RTC binding rate is equal to 1 (\textmu M s)$^{-1}$ and the unbinding rate is 0.1 s$^{-1}$. The translation rates are computed using the length of each protein and considering a constant speed for the ribosome. We study the impact of changing of this parameter (ribosome speed) in Fig. \ref{virion release}a. We observe that by changing ribosome speed the model behaves qualitatively the same. Thus, we fix this parameter in other simulations and we assume that the speed of the ribosome is 10 amino acids per second. The replication and transcription rates of RTC are computed using the length of each RNA and considering a constant speed for RTC \cite{Kim2020}. We assume that the speed of the RTC is 30 nucleotides per second \cite{Arnold2004,Arias2008}, although this parameter has also been varied. See Table \ref{param} for full details on these constants.

\subsubsection{Remdisivir binding rates}
The relative free energy of binding for remdesivir is $\Delta G=-8.28 \pm 0.65 \mbox{ kcal/mol}$ \cite{Zhang2020}. The following equation relates the binding free energy ($\Delta G$) to $K_d$:
$K_d=e^{\beta\Delta G},$ where $\beta=\dfrac{1}{k_BT}$, $k_B$ is the Boltzmann constant and $T$ is Kelvin temperature \cite{Dykeman2013}. Therefore, for remdesivir $K_d$ is approximately 1~\textmu M and we assume that the binding rate is equal to 1 (\textmu M s)$^{-1}$ and the unbinding rate is 1 s$^{-1}$.

The binding rates are measured in (M s)$^{-1}$, but as we are using a reaction based model with the Gillespie algorithm, we use the cellular volume to change these units to (molecule hour)$^{-1}$ \cite{Dykeman2013}. For changing the units of parameters related to the ribosome, we have used the volume from yeast, but for RTC and remdesivir, we have used the volume of an E. coli cell as these parameters were measured in E. coli. The level of drug in each cell is constant and equal to 25 molecules. Since $V=0.7 \mu m^3$, it is equivalent to 0.06 \textmu M \cite{Gordon2020}.

\section{Intercellular model in the context of antiviral therapy}

In order to study the effects of antiviral therapy in the context of our intercellular model, we multiply the viral production rate $p$ by $(1-\epsilon)$, where $0\leq\epsilon\leq 1$ is the drug efficacy \cite{Dahari2009,Kim2012,Guedj2013,Chenar2018}. As our intracellular model (Fig. 1c and d) suggests that starting treatment in the \textit{latent period} is most effective, and would effectively block the production of virions, we set $\gamma=0$ at the onset of treatment. This means that infected cells in the latent phase ($L$) do not transition to phase $I$, and begin shedding virions at a reduced rate (following a time lag $\tau$) compared with cells that were already in phase $I$ at the onset of treatment. Thus, we have the following equations for the numbers of infected cells and free virions;
\begin{equation}\label{intertreatment}
\begin{array}{l}
\vspace{0.15cm}
\displaystyle{\dfrac{dL}{dt}=\beta TV-\delta L-\gamma(1-\theta(t-t_R))L-\mu LE,}\\
\vspace{0.15cm}
\displaystyle{\dfrac{dI}{dt}=\gamma(1-\theta(t-t_R)) L-\delta I-\mu IE,}\\
\displaystyle{\dfrac{dV}{dt}=p(1-\epsilon)\theta(t-\tau-t_R)L(t-\tau)+p(1-\eta\theta(t-t_R))I-cV-kAV,}\\
\end{array}
\end{equation}
where $\theta(.)$ is the Heaviside function ($\theta(t)=1, \mbox{ for } t \geq 0, \mbox{ and } \theta(t)=0, \mbox{ for } t < 0 $), $\epsilon$ and $\eta$ are efficacies of the drug for cells that are in phases $L$ and $I$, respectively, and $t_R$ is the time when the antiviral treatment (remdesivir) is introduced.
\newpage
\section{Structural identifiability of the intercellular model}

We use the method by Castro and de Boer \cite{Castro2020}, that is based on scale invariance of the equations, to establish structural identifiability of our model. Following this method, we scale parameters and variables by unknown scaling factors, denoted here as $u_j$ with indices $j$ used to distinguish them. Assuming otherwise fixed parameters, and given that viral load $V$ is the observable, we have $u_V=1$. We substitute the new rescaled parameters and variables into the model to obtain:

\begin{equation*}
\begin{array}{l}
\vspace{0.15cm}
\displaystyle{\dfrac{dT}{dt}=\dfrac{1}{u_T}(u_Tr_TT(1-\dfrac{u_TT+u_LL+u_LI}{T_m})-u_{\beta}u_T\beta TV),}\\\vspace{0.15cm}
\displaystyle{\dfrac{dL}{dt}=\dfrac{1}{u_L}(u_{\beta}u_T\beta TV-u_{\delta}u_L\delta L-u_{\gamma}u_L\gamma L-u_{\mu}u_Lu_E\mu LE),}\\\vspace{0.15cm}
\displaystyle{\dfrac{dI}{dt}=\dfrac{1}{u_I}(u_{\gamma}u_L\gamma L-u_{\delta}u_I\delta I-u_{\mu}u_Iu_E\mu IE),}\\\vspace{0.15cm}
\displaystyle{\dfrac{dV}{dt}=u_pu_IpI-u_ccV-u_ku_AkAV,}\\\vspace{0.15cm}
\displaystyle{\dfrac{dE}{dt}=\dfrac{1}{u_E}(\lambda_E+u_{\alpha}u_E\alpha(u_LL+u_II)E-u_{d_E}u_Ed_EE),}\\
\displaystyle{\dfrac{dA}{dt}=\dfrac{1}{u_A}(u_{p_A}p_AV+u_{r_A}u_Ar_AA(1-\dfrac{u_AA}{A_m})-u_ku_AkAV-u_Ad_AA).}
\end{array}
\end{equation*}

We then equate the right hand sides of these equations to the right hand sides in the model (2.1), and then solve for the values of these scaling factors. It is easy to show that for all parameters and variables the scaling factor is implied to be equal to 1. Thus, according to the method by Castro and de Boer the model is structurally identifiable.
\color{black}
\newpage
\newpage
\newpage
\newpage
\begin{figure}[H]
	\centering
	\includegraphics[width=0.8\linewidth]{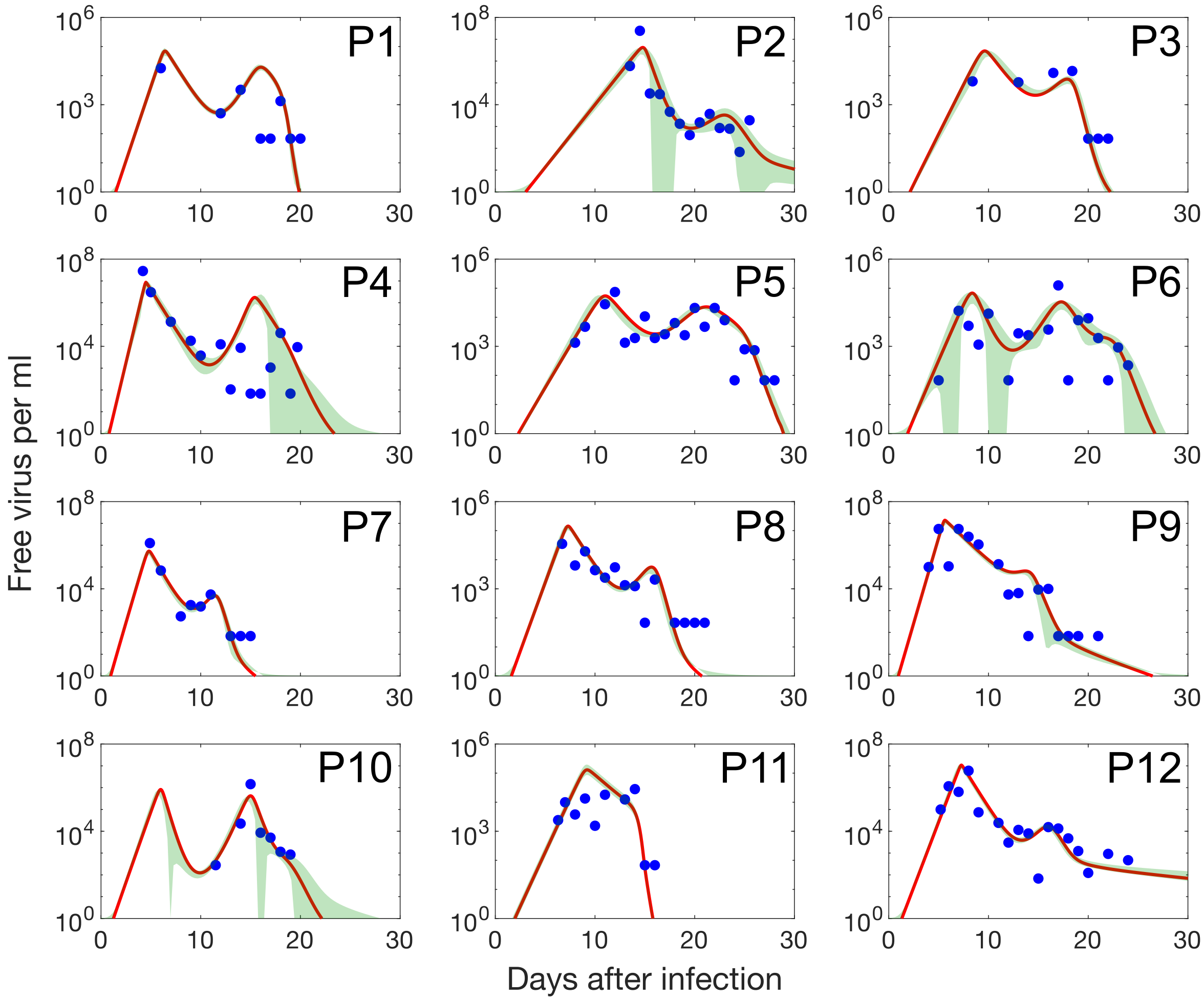}
	\caption{Comparison of patient data with model predictions. The best fit of viral load $V$ computed using the model in (2.1) (red line) to data for 12 patients (\textcolor{blue}{$\bullet $}) is presented. The 95\% CI (the confidence interval) is shown as a red shaded area around each curve, but is indistinguishable from it by eye due it its small size. The green shaded areas indicate the mean plus/minus standard deviation, reflecting the fact that some patient data (such as P4 and P6) are more noisy, resulting in bigger fluctuations in the fitted curves.}
	\label{patient data}
\end{figure}
\newpage
\begin{figure}[H]
	\centering
	\includegraphics[width=1\linewidth]{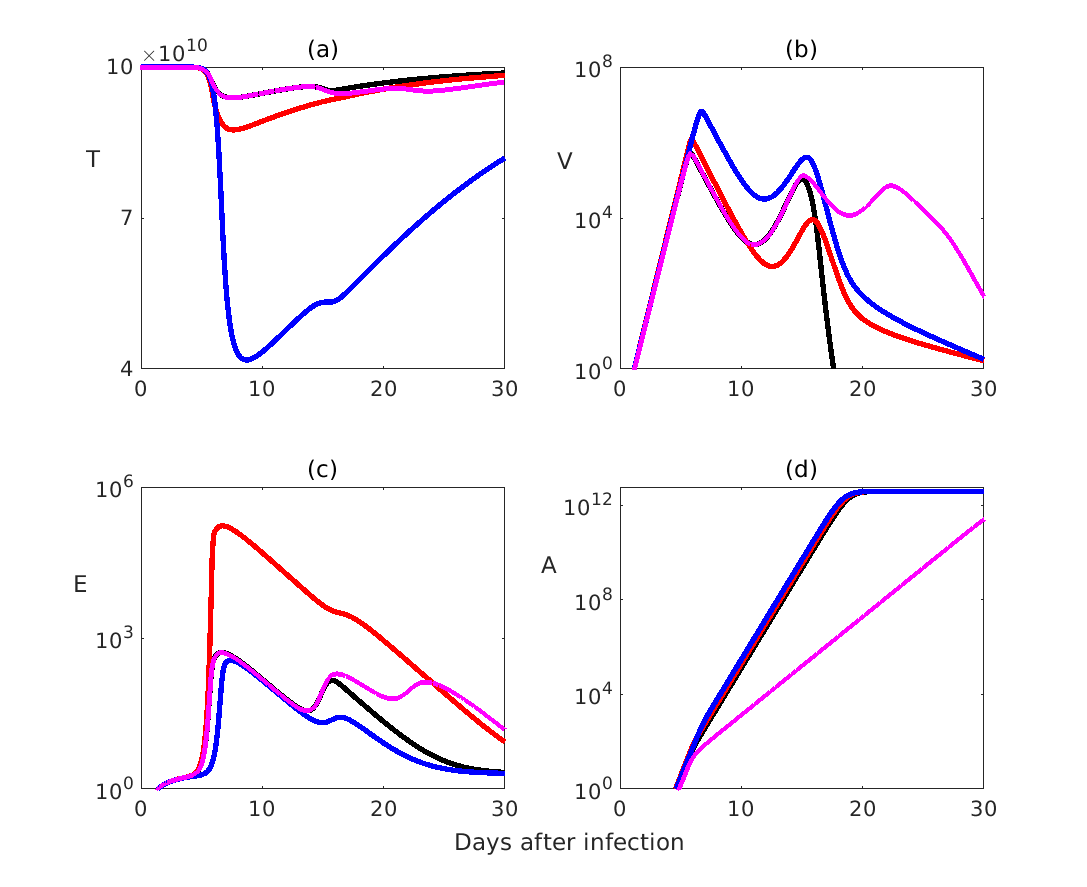}
	\caption{The impact of the immune response on infection dynamics. Parameter values used are the median values of Table 1 (black lines are given by $\mu=5.7\times 10^{-2}$, $\alpha=7.08\times 10^{-9}$ and $r_A=1.98$). A reduction in the removal rate of infected cells by effector cells causes a spike in the level of effector cells (red lines given by $\mu=3.5\times 10^{-4}$). A decrease of the proliferation rate of effector cells causes significant damage to healthy cells and increases the viral peak (blue lines are given by $\alpha=5.4\times 10^{-10}$), and reduction of the proliferation rate of antibodies indicates a third peak in viral load (magenta lines are given by $r_A=1$).}
	\label{immune}
\end{figure}
\newpage
\begin{figure}[H]
	\centering
	\includegraphics[width=1\linewidth]{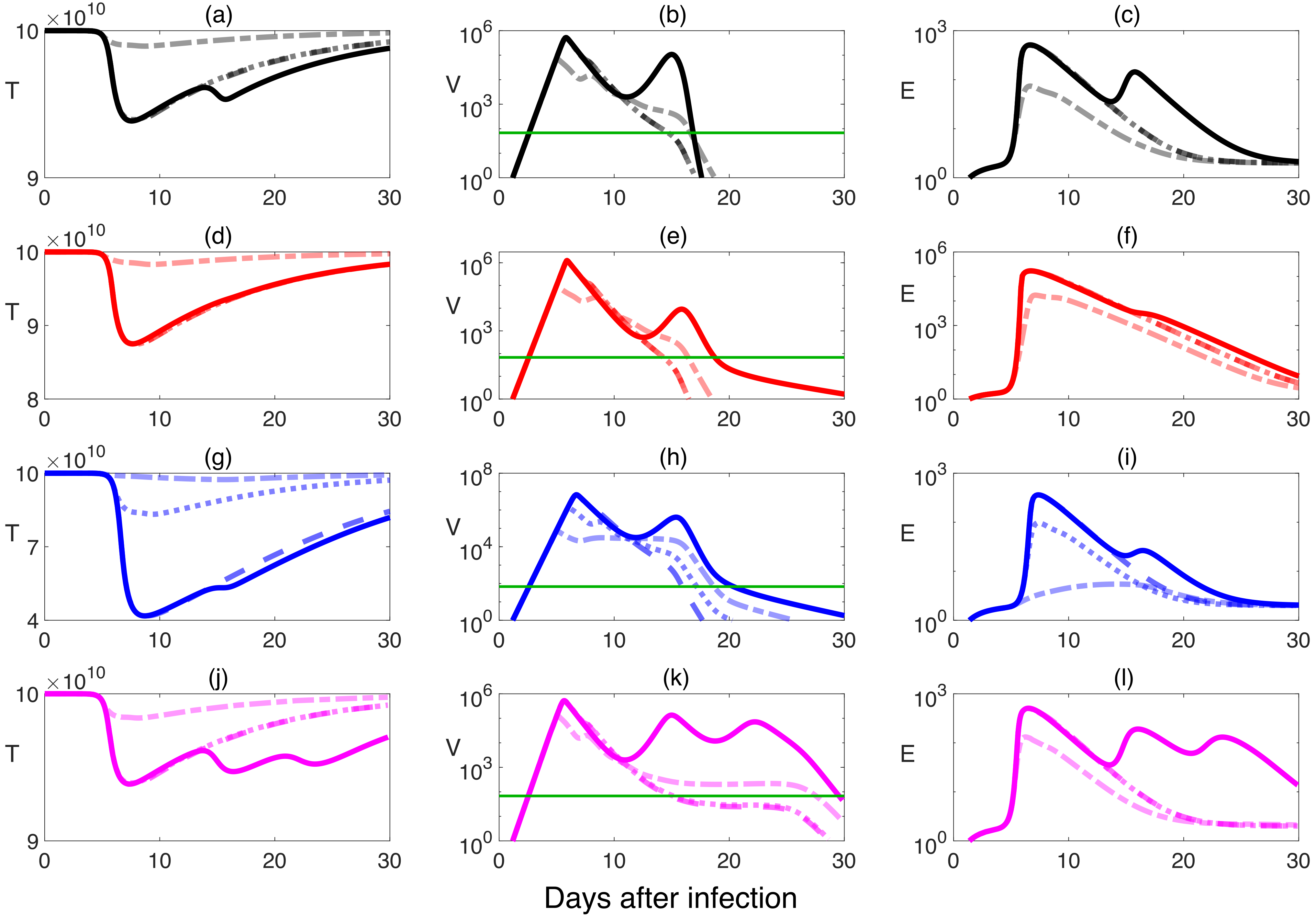}
	\caption{Treatment model with a time lag, i.e. infected cells in group $L$ do not move to $I$ after starting of treatment and produce virions after a time lag $\tau$ (\ref{intertreatment}). Solid lines indicate progression of the infection in the absence of treatment as a control. Dashed, dotted and dash-dotted curves show the result of starting treatment at 7, 6, and 5 dpi, respectively. Parameters are the median values of Table 1 with $\tau=2\mbox{ days}$. Red curves correspond to the scenario of low removal rate of infected cells by effector cells ($\mu=3.5\times 10^{-4}$). Blue curves illustrate the scenario of a low proliferation rate of effector cells ($\alpha=5.4\times 10^{-10}$), and magenta curves of a low antibody proliferation rate ($r_A=1$). The green line indicates the viral detection limit. Outcomes are similar to the model in the absence of time lag, which is therefore used in the main text.}
	\label{treatment delay}
\end{figure}
\newpage
\begin{figure}[H]
	\centering
	\includegraphics[width=\linewidth]{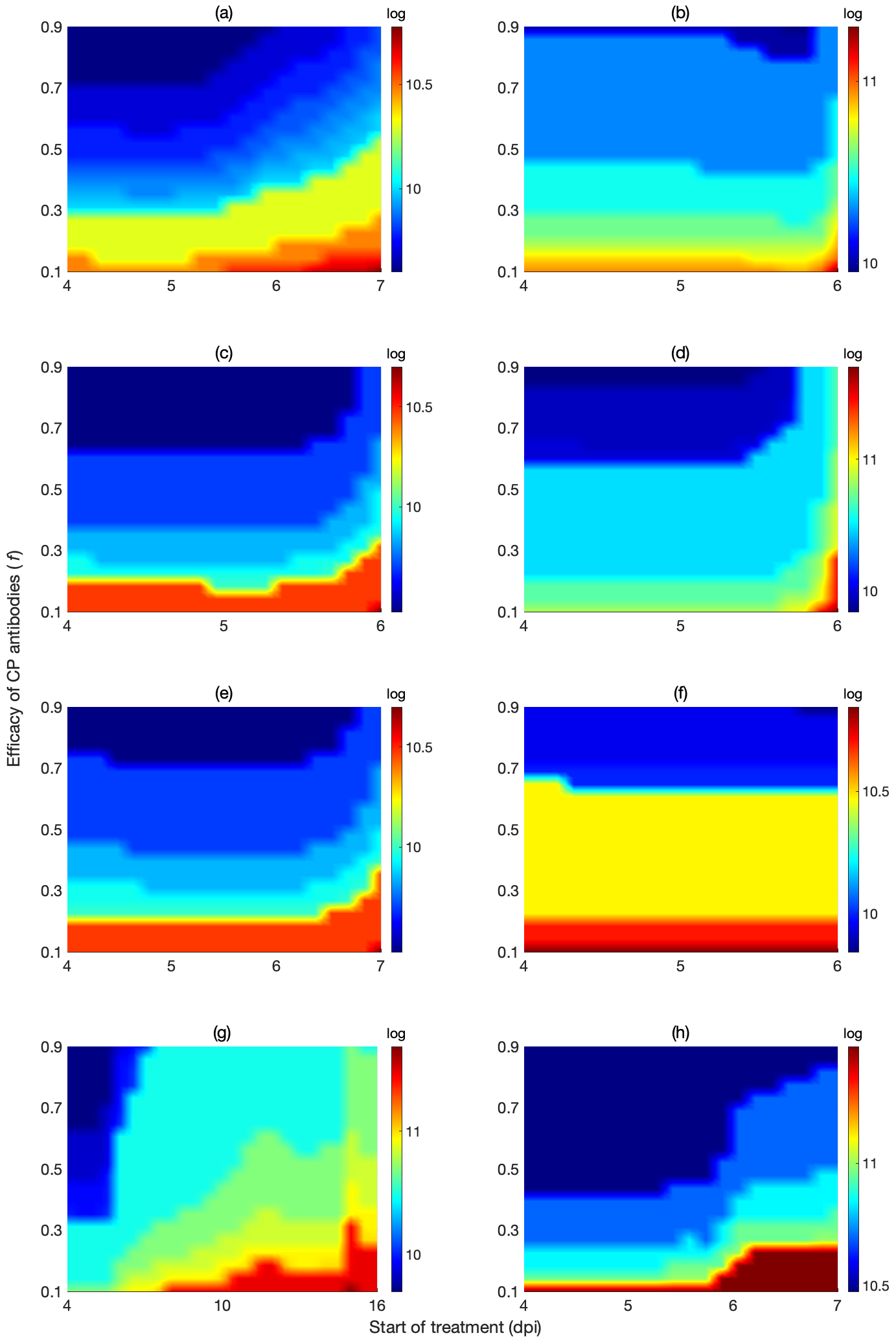}
	\caption{The impact of CP therapy on the reduction of AUC for different treatment starts and antibody efficacy. The first and second column illustrate the minimum level of $\tilde{A}_m$ required for 25\% and 50\% reduction in the AUC, respectively. (a) and (b) Parameters are the median values of Table 1, i.e. represent the generic case based on the patient data; (c) and (d) correspond to the scenario of low removal rate of infected cells by effector cells ($\mu=3.5\times 10^{-4}$); (e) and (f) illustrate the scenario of a low proliferation rate of effector cells ($\alpha=5.4\times 10^{-10}$); and (g) and (h) show a case with a low antibody proliferation rate ($r_A=1$).} 
	\label{surf}
\end{figure}
\newpage
\begin{figure}[H]
	\centering
	\includegraphics[width=\linewidth]{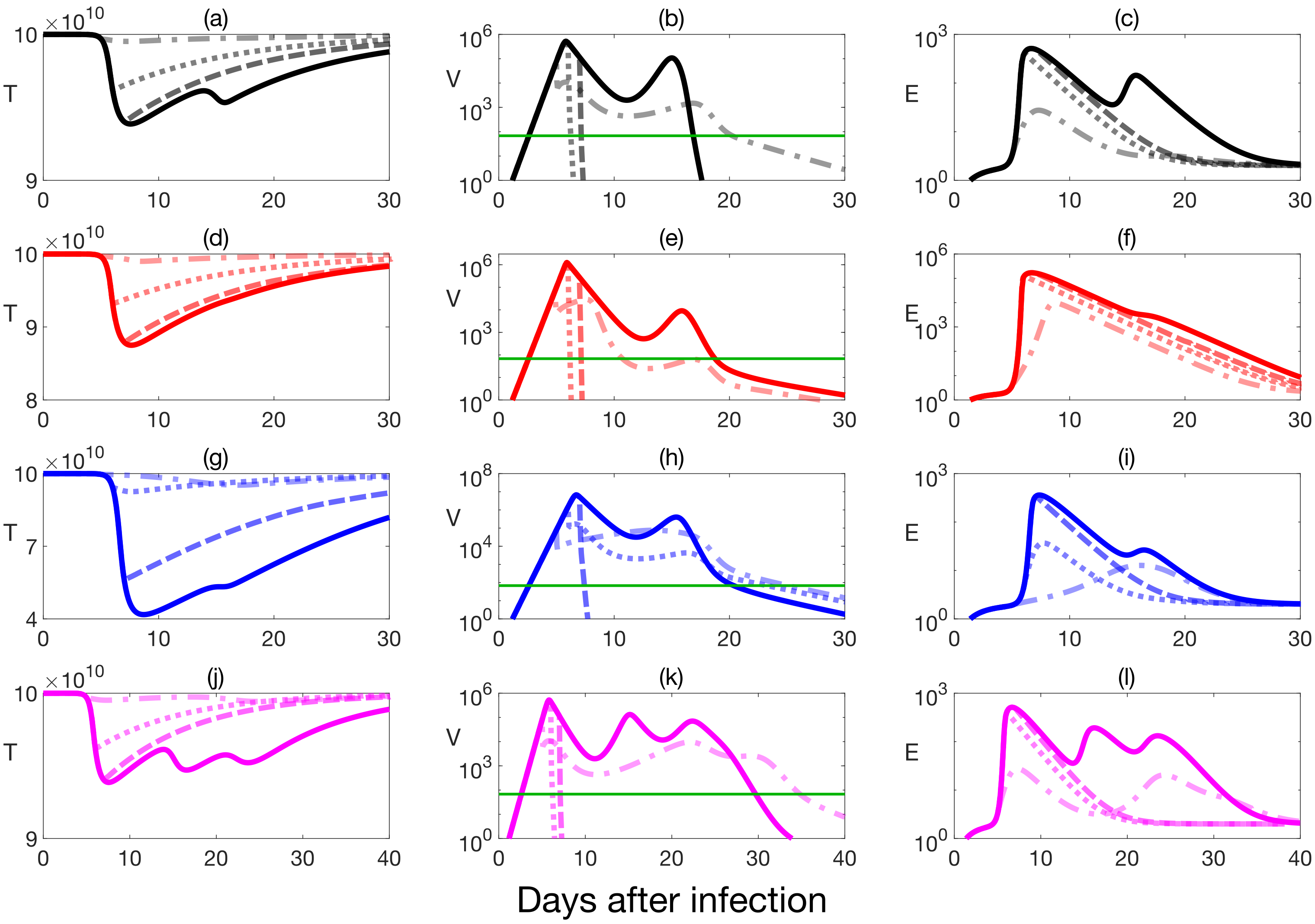}
	\caption{An early CP therapy increases the duration of infection more compared with an early antiviral therapy.  Solid lines indicate progression of the infection in the absence of treatment as a control. Dashed, dotted and dash-dotted curves show the result of starting treatment at 7, 6, and 5 dpi, respectively.  Parameters are the median values of Table 1. Red curves correspond to the scenario of low removal rate of infected cells by effector cells ($\mu=3.5\times 10^{-4}$). Blue curves illustrate the scenario of a low proliferation rate of effector cells ($\alpha=5.4\times 10^{-10}$), and magenta curves of a low antibody proliferation rate ($r_A=1$). The green line indicates the viral detection limit.}
	\label{mAb treatment}
\end{figure}
\newpage
\begin{table}[H]
\centering
\caption{Table of parameter values}
\begin{tabular}{c|c|c|c}
\hline\hline
 Parameter & Value & Parameter & Value \\\hline
 $r_{on}$ & 8.64 Molecule$^{-1}$hour$^{-1}$ & $r_{off}$ & 7200 hour$^{-1}$ \\
 $d_r$ & 0.11 hour$^{-1}$ & $r_{in}$ & 360 hour$^{-1}$ \\
 $t_{1a}$ & 8.2 hour$^{-1}$ & $t_{1b}$ & 13.4 hour$^{-1}$ \\
 $t_f$ & 360 hour$^{-1}$ & $q$ & 0.4 \\
 $\delta_a$ & 0.69 hour$^{-1}$ & $\delta_b$ & 0.69 hour$^{-1}$ \\
 $f_{rt}$ & 30 hour$^{-1}$ & $\delta_r$ & 0.11 hour$^{-1}$ \\
 $rt_{on}$ & 8.64 Molecule$^{-1}$hour$^{-1}$ & $rt_{off}$ & 360 hour$^{-1}$ \\
 $rt_{in}$ & 360 hour$^{-1}$ & $r$ & 0.7 \\
 $t_n$ & 4.3 hour$^{-1}$ & $t_m$ & 8.1 hour$^{-1}$ \\
 $t_e$ & 24 hour$^{-1}$ & $t_s$ & 1.4 hour$^{-1}$ \\
 $\delta_n$ & 0.69 hour$^{-1}$ & $\delta_m$ & 0.69 hour$^{-1}$ \\
 $\delta_e$ & 0.69 hour$^{-1}$ & $\delta_s$ & 0.69 hour$^{-1}$ \\
 $t_c$ & 360 hour$^{-1}$ & $r_{term}$ & 360 hour$^{-1}$ \\
 $k_{bud}$ & 0.0001 hour$^{-1}$ & $Rib$ & 1000 \\
 $k_{on}$ & 8.64 Molecule$^{-1}$hour$^{-1}$ & $k_{off}$ & 3600 hour$^{-1}$ \\\hline
 \textbf{tr} & \multicolumn{3}{l}{(66.26, 284.21, 138d0, 782.61, 562.5, 159.1, 388.49, 126.76, 28.2, 5) hour$^{-1}$}\\
 \textbf{b} & \multicolumn{3}{l}{(8.64, 8.64, 8.64, 8.64, 8.64, 8.64, 8.64, 8.64, 8.64, 8.64) hour$^{-1}$}\\
 \textbf{d} & \multicolumn{3}{l}{(0.11, 0.11, 0.11, 0.11, 0.11, 0.11, 0.11, 0.11, 0.11, 0.11) hour$^{-1}$}\\
 \textbf{n} & \multicolumn{3}{l}{(1, 1, 1, 1, 1, 1, 1, 1, 1, 1)} \\
 \textbf{p} & \multicolumn{3}{l}{(63.53, 51.92, 48.69, 41.86, 38.96, 31.29, 28.96, 23.57, 12.84, 3.61) hour$^{-1}$}\\\hline
\end{tabular}
\label{param}
\end{table}

\begin{sidewaystable}
\footnotesize
\centering
\caption{Bootstrap 95\% CIs for the parameter estimates}
\label{patient param 95CI}
\begin{tabular}{lcccccccccccc}
\hline
Patient & $\beta\times 10^{-8}$ & $\delta$ & $\gamma$ & $\mu$   & $p\times 10^{-3}$ & $k\times 10^{-10}$ & $c$   & $\alpha\times 10^{-9}$ & $p_A\times 10^{-5}$ & $r_A$ & I.P & A.A \\\hline
1  & {[}20.685, 20.715{]} & {[}0.247, 0.249{]}     & {[}0.899, 0.901{]}       & {[}0.0464, 0.0476{]}       & {[}2.153, 2.207{]}   & {[}2.338, 2.402{]} & {[}1.236, 1.244{]}   & {[}17.162, 17.238{]} & {[}1.288, 1.372{]} & {[}1.976, 1.984{]} & {[}3.97, 4.03{]}   & {[}14.97, 15.03{]}   \\
2  & {[}2.522, 2.558{]}   & {[}0.246, 0.25{]}      & {[}0.899, 0.901{]}       & {[}0.0052, 0.0064{]}       & {[}8.073, 8.187{]}   & {[}3.65, 3.71{]}   & {[}3.11, 3.19{]}     & {[}1.149, 1.191{]}   & {[}0.842, 0.898{]} & {[}1.624, 1.636{]} & {[}3.95, 4.05{]}   & {[}20.95, 21.05{]}   \\
3  & {[}9.02, 9.22{]}     & {[}0.341, 0.347{]}     & {[}0.898, 0.902{]}       & {[}0.096, 0.104{]}         & {[}2.087, 2.153{]}   & {[}2.609, 2.771{]} & {[}1.16, 1.2{]}      & {[}15.027, 15.973{]} & {[}1.362, 1.458{]} & {[}2.045, 2.055{]} & {[}5.98, 6.02{]}   & {[}15.98, 16.02{]}   \\
4  & {[}16.09, 16.91{]}   & {[}0.194, 0.206{]}     & {[}0.99996, 1.00004{]}   & {[}0.00098, 0.00102{]}     & {[}9.267, 9.533{]}   & {[}2.116, 2.284{]} & {[}1.57, 1.65{]}     & {[}1.257, 1.483{]}   & {[}4.267, 4.593{]} & {[}1.233, 1.267{]} & {[}1.999, 2.001{]} & {[}15.999, 16.001{]} \\
5  & {[}5.808, 5.972{]}   & {[}0.168, 0.172{]}     & {[}0.999994, 1.000006{]} & {[}0.112, 0.118{]}         & {[}1.929, 1.991{]}   & {[}5.52, 5.58{]}   & {[}1.12, 1.16{]}     & {[}19.045, 19.155{]} & {[}1.879, 2.001{]} & {[}1.368, 1.372{]} & {[}5.96, 6.04{]}   & {[}22.96, 23.04{]}   \\
6  & {[}12.79, 13.21{]}   & {[}0.268, 0.282{]}     & {[}0.818, 0.822{]}       & {[}0.182, 0.192{]}         & {[}3.116, 3.284{]}   & {[}4.883, 5.117{]} & {[}1.82, 1.96{]}     & {[}7.364, 8.416{]}   & {[}2.969, 3.231{]} & {[}1.47, 1.49{]}   & {[}1.998, 2.002{]} & {[}19.998, 20.002{]} \\
7  & {[}13.15, 13.45{]}   & {[}0.193, 0.207{]}     & {[}0.847, 0.853{]}       & {[}0.19998, 0.20002        & {[}9.803, 9.997{]}   & {[}4.156, 4.404{]} & {[}1.94, 2{]}        & {[}5.927, 6.613{]}   & {[}6.457, 6.743{]} & {[}2.822, 2.838{]} & {[}3.97, 4.03{]}   & {[}9.97, 10.03{]}    \\
8  & {[}10.29, 10.51{]}   & {[}0.346, 0.354{]}     & {[}0.79998, 0.80002{]}   & {[}0.038, 0.042{]}         & {[}3.587, 3.713{]}   & {[}3.408, 3.592{]} & {[}1.28, 1.32{]}     & {[}16.275, 16.725{]} & {[}6.923, 7.077{]} & {[}2.173, 2.187{]} & {[}3.96, 4.04{]}   & {[}13.96, 14.04{]}   \\
9  & {[}10.43, 10.77{]}   & {[}0.50995, 0.51005{]} & {[}0.837, 0.843{]}       & {[}0.0029, 0.0041{]}       & {[}10.336, 10.464{]} & {[}2.713, 2.887{]} & {[}1.09, 1.11{]}     & {[}0.829, 0.891{]}   & {[}0.938, 0.982{]} & {[}2.124, 2.136{]} & {[}2.98, 3.02{]}   & {[}12.98, 13.02{]}   \\
10 & {[}12.05, 12.35{]}   & {[}0.219, 0.239{]}     & {[}0.79997, 0.80003{]}   & {[}0.114, 0.126{]}         & {[}10.077, 10.323{]} & {[}2.533, 2.787{]} & {[}3.89, 3.93{]}     & {[}2.9, 3.1{]}       & {[}5.562, 5.738{]} & {[}1.564, 1.576{]} & {[}5.96, 6.04{]}   & {[}15.96, 16.04{]}   \\
11 & {[}12.471, 12.529{]} & {[}0.196, 0.204{]}     & {[}0.896, 0.904{]}       & {[}0.0664, 0.0676{]}       & {[}1.275, 1.325{]}   & {[}3.952, 4.048{]} & {[}0.7496, 0.7504{]} & {[}7.911, 8.269{]}   & {[}6.415, 6.585{]} & {[}2.491, 2.509{]} & {[}5.98, 6.02{]}   & {[}12.98, 13.02{]}   \\
12 & {[}6.036, 6.104{]}   & {[}0.196, 0.204{]}     & {[}0.898, 0.902{]}       & {[}0.0009998, 0.0010002{]} & {[}9.267, 9.353{]}   & {[}1.237, 1.283{]} & {[}1.68, 1.72{]}     & {[}1.103, 1.137{]}   & {[}7.615, 7.785{]} & {[}1.975, 1.985{]} & {[}3.97, 4.03{]}   & {[}13.97, 14.03{]}   \\\hline
\end{tabular}
\footnotetext{I.P: incubation period (day). A.A: antibodies appearance (day).}
\footnotetext{The units of these parameters are day$^{-1}$.}
\end{sidewaystable}

\newpage

\begin{table}[H]
\color{black}
\caption{AUC values for antiviral therapy}
\begin{tabular}{lllll}
        & No Treatment      & 5 dpi             & 6 dpi            & 7 dpi             \\ \hline
Healthy I.S (median values, Fig. 3b) & $8.51\times 10^5$ & $1.22\times 10^5$ & $6.6\times 10^5$ & $6.73\times 10^5$ \\ \hline
Immunocompromised (low $\mu$, Fig. 3e) & $1.43\times 10^6$ & $1.95\times 10^5$ & $1.4\times 10^6$ & $1.41\times 10^6$ \\ \hline
Immunocompromised (low $\alpha$, Fig. 3h) & $9.81\times 10^6$ & $4.37\times 10^5$ & $2.5\times 10^6$ & $8.77\times 10^6$ \\ \hline
Immunocompromised (low $r_A$, Fig. 3k) & $1.23\times 10^6$ & $1.23\times 10^5$ & $6.6\times 10^5$ & $6.73\times 10^5$ \\ \hline
\end{tabular}\\
\footnotesize{I.S: immune system}
\end{table}

\begin{table}[H]
\caption{AUC values for CP therapy}
\begin{tabular}{lllll}
        & No Treatment      & 5 dpi             & 6 dpi            & 7 dpi             \\ \hline
Healthy I.S (median values, Fig. 3b) & $8.51\times 10^5$ & $5.78\times 10^4$ & $3.51\times 10^5$ & $5.94\times 10^5$ \\ \hline
Immunocompromised (low $\mu$, Fig. 3e) & $1.43\times 10^6$ & $1.03\times 10^5$ & $6.29\times 10^5$ & $1.23\times 10^6$ \\ \hline
Immunocompromised (low $\alpha$, Fig. 3h) & $9.81\times 10^6$ & $6.22\times 10^5$ & $7.93\times 10^5$ & $5.2\times 10^6$ \\ \hline
Immunocompromised (low $r_A$, Fig. 3k) & $1.23\times 10^6$ & $1.22\times 10^5$ & $3.51\times 10^5$ & $5.94\times 10^5$ \\ \hline
\end{tabular}\\
\footnotesize{I.S: immune system}
\end{table}

\begin{table}[H]
\caption{AUC values for antiviral+CP therapy}
\begin{tabular}{lllll}
        & No Treatment      & 5 dpi             & 6 dpi            & 7 dpi             \\ \hline
Healthy I.S (median values, Fig. 3b) & $8.51\times 10^5$ & $2.9\times 10^4$ & $3.51\times 10^5$ & $5.94\times 10^5$ \\ \hline
Immunocompromised (low $\mu$, Fig. 3e) & $1.43\times 10^6$ & $3.84\times 10^4$ & $6.29\times 10^5$ & $1.23\times 10^6$ \\ \hline
Immunocompromised (low $\alpha$, Fig. 3h) & $9.81\times 10^6$ & $3.03\times 10^4$ & $5.37\times 10^5$ & $5.2\times 10^6$ \\ \hline
Immunocompromised (low $r_A$, Fig. 3k) & $1.23\times 10^6$ & $2.9\times 10^4$ & $3.51\times 10^5$ & $5.94\times 10^5$ \\ \hline
\end{tabular}\\
\footnotesize{I.S: immune system}
\end{table}

\begin{table}[H]
\caption{AUC values for treatment starts after the reported onset of symptoms}
\begin{tabular}{lllll}
       & No treatment        & Antiviral         & CP                & Antiviral+CP      \\\hline
P1     & $1.85\times 10^{5}$ & 5420              & $1.17\times 10^4$ & 365.08            \\\hline
P2     & $6.13\times 10^6$   & 132.08            & 226.72            & 51.35             \\\hline
P3     & $1.77\times 10^5$   & 1200              & 512.21            & 312.85            \\\hline
P4     & $1.11\times 10^7$   & $4.66\times 10^5$ & $1.28\times 10^6$ & 77.36             \\\hline
P5     & $2.41\times 10^5$   & 304.46            & 102.16            & 84.03             \\\hline
P6     & $3.41\times 10^5$   & 204.63            & 84.89             & 53.77             \\\hline
P7     & $5.9\times 10^5$    & $8.08\times 10^4$ & $3.76\times 10^4$ & $1.76\times 10^4$ \\\hline
P8     & $2.41\times 10^5$   & 766.43            & 363.77            & 90.23             \\\hline
P9     & $1.79\times 10^7$   & $1.25\times 10^5$ & $8.21\times 10^5$ & 445.64            \\\hline
P10    & $1.26\times 10^6$   & $6.38\times 10^5$ & $4.34\times 10^5$ & $4.28\times 10^5$ \\\hline
P11    & $3.24\times 10^5$   & 4500.1            & 1008.6            & 629.8             \\\hline
P12    & $1.19\times 10^7$   & $2.19\times 10^4$ & $1.71\times 10^6$ & 888.37            \\\specialrule{.2em}{.1em}{.1em} 
median & $4.66\times 10^5$   & 4960              & 6350              & 339               \\\hline
mean   & $4.2\times 10^6$    & $1.12\times 10^5$ & $3.58\times 10^5$ & $3.74\times 10^4$ \\\hline
std    & $6.13\times 10^6$   & $2.1\times 10^5$ & $5.95\times 10^5$ & $1.23\times 10^5$ \\\hline
95\% CI& $[0.73, 7.7]\times 10^6$   & $[0.00001, 2.3]\times 10^5$ & $[0.21, 6.95]\times 10^5$ & $[0.000001, 1.07]\times 10^5$ \\\hline
\end{tabular}
\end{table}
\newpage